# Mathematical Properties of Incremental Effect Additivity and Other Synergy Theories

Leonid Hanin[1*], Liyang Xie[2], Rainer Sachs[2]

[1] Department of Mathematics and Statistics, Idaho State University, 921 S. 8th Avenue, Stop 8085, Pocatello, ID 83209-8085, USA

[2] Department of Mathematics, University of California, Berkeley, 970 Evans Hall, Berkeley, CA 94720-3840, USA

*Corresponding author. E-mail: hanin@isu.edu

**Abstract** Synergy theories for multi-component agent mixtures use 1-agent dose-effect relations (DERs), assumed known from analyzing previous 1-agent experiments, to calculate baseline Neither-Synergy-Nor-Antagonism (NSNA) mixture DERs. The most commonly used synergy theory, Simple Effect Additivity (SEA), is not self-consistent mathematically. Many nonlinear alternatives have been suggested, almost all of which require an assumption that effects increase monotonically as dose increases. We here emphasize the recently introduced Incremental Effect Additivity (IEA) synergy theory and briefly discuss Loewe additivity. By utilizing the fact that, when dose increments approach zero, dose-effect relations approach linearity, IEA theory to some extent circumvents the DER nonlinearities that plague SEA calculations. We study mathematical properties of IEA that are relevant to practical implementation of this synergy theory and hold whatever particular area of biology, medicine, toxicology or pharmacology is involved. However, as yet IEA has only been applied to mixture experiments simulating the toxic galactic cosmic ray mixture encountered during voyages in interplanetary space. Our main results are theorems, propositions, examples and counterexamples revealing various properties of IAE, including whether or not IEA NSNA DERs lie between mixture components' DERs. These results are amply illustrated with figures.

# 1 Introduction

## 1.1 Synergy Theories

When several agents act jointly on a biological or other object a question of great theoretical and practical importance is how to detect the presence of synergistic or antagonistic interactions between the agents based on measurements of the effect produced by various doses of these agents in experiments where the agents act either separately or in different combinations. In this article, we assume that (1) the agents have a common target, (2) their doses can be measured using the same units, and (3) the agents produce an observable effect that can be measured on the same scale.

A key idea behind the detection of interaction between several agents is to specify, for every mixture dose $d$ of these agents, a certain value, $I(d)$, of effect that characterizes the lack of either synergy or antagonism between the agents. If the observed joint effect is larger that $I(d)$ then the action of the agents is assumed synergistic while if it is smaller than $I(d)$ then the combined action of the agents is viewed as antagonistic. For a review of this approach, see Berenbaum (1989) and Huang et al. (2020). In this article, the function $I(d)$ will be called *Neither-Synergy-Nor-Antagonism (NSNA) baseline*; another commonly used term is *null reference model*, see e.g. Lederer et al. 2018. A specific way of defining and justifying the function $I(d)$ will be referred to as *synergy theory*. Analyzing statistical significance of synergy or antagonism based on empirical data is beyond the scope of this work.

The main goal of this article is to systematically develop and study mathematical properties of *Incremental Effect Additivity* (IEA), a major synergy theory introduced recently in Ham et al. 2018. In our opinion, IEA formalism can be used in many different fields including radiation biology, microbiology, toxicology, pharmacology and epidemiology. One notable application, see Huang et al. (2020), is to simulations of biological effects of the multi-component galactic cosmic ray mixture in interplanetary space (Norbury et al. 2019). This particular application colors some of our methodological choices and terminology. For example, we will usually regard dose $d$ as a variable which only allowed transformations correspond to a change in units, e.g. from gray (Gy) to milligray (mGy), i.e. scaling. In radiobiology, this stringent restriction of allowed dose transformations is natural because dose has a fundamental physical meaning – mean energy absorbed per unit mass – that cannot conveniently be ignored. Another consequence of the above-mentioned application is an emphasis on synergy theories that can readily handle mixtures of many components rather than just two.

Throughout this paper we emphasize deleterious effects and assume, in intuitive interpretations, that agents are harmful. We also intuitively interpret synergy as "bad" synergy – even more damage than expected from non-interacting mixture components. By changing heuristic interpretations, the formalisms we discuss could instead be applied to situations where agents intended to cure are involved, as is common in pharmacology.

In this article, we only consider situations where dose is administered acutely – i.e. so rapidly it can be regarded as almost instantaneous. Alternatives – including the case, very important in some applications, of a low, approximately constant dose rate – are not discussed.

**1.2 Dose, Effect, Dose-Effect Relations (DERs) and Effect-Dose Relations (EDRs)**

Let $E(d)$ be a function that associates with every admissible dose $d$ of a given agent the effect, $E(d)$, produced by the agent in a biological or other object on which it acts. Such functions, called *Dose-Effect Relations* (DERs), play a key role in most synergy theories. These theories use 1-agent DERs, assumed already known due to analyzing the results of experiments where agents in a mixture act alone, to derive a Neither-Synergy-Nor-Antagonism (NSNA) DER, $I(d)$, for a mixture of those agents.

Every 1-agent DER $E(d)$ is required to have domain *[0, ∞)* and *E(0) = 0* must hold. Our DERs thus always have background subtracted out, for our intention is to study synergy amongst mixture agents, not synergy between known agents and unknown causes of effects in untreated, control targets. We regard 1-agent DERs as informed estimates based on analyzing 1-agent data but not as random functions accounting for variation of data used to produce statistical estimates of individual DERs. Selecting 1-agent DERs involves assembling information from many different sources – studying 1-agent data visually, employing heterogeneous information on a specific biologal endpoint, using published calculations, analyzing biological mechanisms, making comparisons among competing DERs, etc.

Consider a mixture of $N > 1$ agents. DER, $E_j$, of $j$-th agent in the mixture is a function $E_j: [0, \infty) \to [0, h_j)$, where $h_j \in (0, \infty]$ is the maximum effect produced by this agent, $1 \leq j \leq N$. Suppose that for every $j$, the $j$-th mixture component contributes dose $d_j$ which constitutes a fraction $r_j = d_j/d > 0$ of the total mixture dose $d$, so that $d_j = r_j d$. Note that $\sum_{j=1}^{N} d_j = d$ or equivalently $\sum_{j=1}^{N} r_j = 1$. For convenience and without essential loss of generality, we assume throughout this paper that each $r_j$ is independent of $d$, which is the most common situation in experiments.

If all effects in a given mixture experiment are subjected to a continuous monotonically increasing transformation that leaves zero invariant, the resulting calculations will usually be considered to describe the same biological situation as the calculations that would have occurred if the transformation had not been made. Thus, allowed transformations of the effect are here less stringently restricted than those for the dose.

Some examples of 1-agent DER $E(d)$ are $d^a$ with $a > 0$; $exp(d) - 1$; $d exp(d)$; $ln(d + 1)$; $d/(d+1)$; a more general Hill model $E(d) = d^a/(d^a+b)$ with $a, b > 0$ that is widely used in pharmacology, genomics and other fields for describing kinetics of biochemical reactions (see e.g. Goutelle et al. 2008); and the linear-quadratic (LQ) function

$$E(d) = \alpha d + \beta d^2, \tag{1.2.1}$$

where α and β are non-negative numbers, with α + β > 0. LQ functions have played a prominent role in radiobiology, including synergy modeling, see e.g. Zaider and Rossi (1980). A more sophisticated example of DER, that is nonetheless accommodated by our IEA theory, is exp($-d^{-a}$) for $a > 0$; every such DER vanishes at $d = 0$ together with all its derivatives and is thus "extremely flat" at the origin. Notice that the range of the above DERs is an interval *[0, h)*, where $h = 1$ or $\infty$.

As reviewed in Huang et al. (2020), there are substantial reasons for sometimes using effect rather than dose as the basic independent variable. In an acutely treated biological target, effects typically persist much longer than dosing – a hint that effects are perhaps more important. If a DER *E(d)* is continuous and monotonic increasing it has an inverse function *F(E)*, the *Effect-Dose Relation* (EDR), and we will often use EDRs in our calculations.

### 1.3 Older Synergy Theories

*1.3.1 Simple Effect Additivity (SEA) Synergy Theory*

Researchers in pharmacology and toxicology have known for a very long time (see Fraser 1872) that the "obvious" method of analyzing mixture effects with the SEA approach to synergy – namely just adding component effects – is unreliable when some mixture components' 1-agent DERs are highly nonlinear. This problem is reviewed, e.g., in Foucquier and Guedj (2015) and Huang et al. (2020).

The NSNA baseline DER that characterizes SEA synergy theory is

$$I(d) = \sum_{j=1}^{N} E_j(r_j d). \tag{1.3.1}$$

A fundamental problem with SEA and many other synergy theories is that they fail to obey what is called the *sham mixture principle*, reviewed in Berenbaum (1989) and Huang et al. (2020), thus making it not self-consistent. For a synergy theory to obey the sham mixture principle, the theory's baseline NSNA mixture DER must always, for a mixture of one agent with itself, give the correct answer. The following example shows that SEA synergy theory does not obey the sham mixture principle unless all 1-agent DERs are linear. Suppose an agent has an LQ DER $E(d) = \alpha d + \beta d^2$ with $\beta > 0$. Regard the agent as a 1:1 mixture of two components, each of which happens to have the same 1-agent DER. Then calculating the baseline NSNA DER for this sham mixture by SEA synergy theory gives $2[\alpha(d/2) + \beta(d/2)^2] = \alpha d + \beta d^2/2$. But, of course, one cannot decrease the beam's toxicity by mental gymnastics, so the correct value is $\alpha d + \beta d^2$.

SEA synergy theory also fails to possess a more general "*mixture of mixtures*" property, which constitutes a more stringent litmus test for a valid synergy theory. This property addresses the situation where some of the agents are mixtures of other agents. Such collections of agents constituting the original agent may overlap or contain identical agents. The mixture of mixtures property posits that a joint action theory expressed in terms of the new collection of agents must agree with the original one under the usual rule for transformation of their weights.

*1.3.2 Loewe Additivity*

There are many suggested replacements for SEA. Oddly, many of them also fail to obey the sham mixture principle. One elegant replacement that does obey the principle and, more generally, has the mixture of mixtures property, is Loewe additivity (Loewe and Muischnek 1926; Loewe 1928). Stated in modern mathematical terms and applied to any number of agents, this major synergy theory defines the NSNA mixture dose $F(E)$ for a given effect $E > 0$ as the appropriately weighted harmonic mean of the 1-agent doses $F_j(E)$, $1 \leq j \leq N$, required to produce the same effect:

$$\frac{1}{F(E)} = \sum_{j=1}^{N} \frac{r_j}{F_j(E)}, \qquad (1.3.2)$$

where $F_j: [0, \infty) \to [0, h)$ is the $j$-th agent's EDR. The validity of the mixture of mixtures property follows directly from Eq. (1.3.2).

If a mixture experiment shows that a mixture dose smaller (or larger) than the dose $d = F(E)$ defined by Eq. (1.3.2) suffices to achieve a given effect $E$, the experimental point is in the synergy (or antagonism) region of the EDR, as shown schematically in Fig. 1. Applications of Loewe additivity synergy theory are reviewed, e.g., in Berenbaum (1989) and Lederer et al. (2018). Loewe additivity numerical calculations are facilitated by the CRAN package LoeweAdditivity that accompanies the article by Azasi et al. (2020).

The properties of the mixture EDR based on Loewe additivity principle encapsulated by Eq. (1.3.2) and the underlying assumptions are stated in the following proposition.

***Proposition 1.3.1.*** *Suppose individual DERs $E_j: [0, \infty) \to [0, h)$ are continuous and monotonic increasing functions with the same finite or infinite range $[0, h)$ and satisfy the condition $E_j(0) = 0$, $1 \leq j \leq N$. Then Eq. (1.3.2) determines a unique continuous and monotonic increasing mixture EDR $F: [0, h) \to [0, \infty)$ such that*

$$min\{F_j(E): 1 \leq j \leq N\} \leq F(E) \leq max\{F_j(E): 1 \leq j \leq N\} \text{ for all } E \in [0, h). \qquad (1.3.3)$$

*Either equality holds for all E iff all N DER functions $E_j$ are identical.*

*Proof.* It follows from Eq. (1.3.2) that function $F$ is well-defined, continuous and monotonic increasing and that $F(0) = 0$. For any weights $r_j > 0$, $1 \leq j \leq N$, such that $\sum_{j=1}^{N} r_j = 1$, the weighted harmonic mean of a vector $\mathbf{x} = (x_1, x_2, \ldots, x_N)$ with positive components is defined by

$$H(\mathbf{x}, \mathbf{r}) = \left( \sum_{j=1}^{N} \frac{r_j}{x_j} \right)^{-1}.$$

Clearly, it satisfies the inequalities

$$min\{x_j: 1 \leq j \leq N\} \leq H(\mathbf{x}, \mathbf{r}) \leq max\{x_j: 1 \leq j \leq N\}, \qquad (1.3.4)$$

where both inequalities are strict unless all components of vector $\mathbf{x}$ are equal. This proves Eq. (1.3.3) and the statement about equality for $E > 0$. The same is also true for $E = 0$. The proof is completed.

Inequality in Eq. (1.3.3) and an equivalent inequality

$$\min\{E_j(d): 1 \leq j \leq N\} \leq E(d) \leq \max\{E_j(d): 1 \leq j \leq N\} \text{ for all } d \geq 0,$$

where $E = F^{-1}$ is the mixture DER, constitute the *betweenness property*. This property is an important advantage of Loewe additivity over other synergy theories. The assumption that all 1-agent DERs must have the same range can be lifted, see subsection 3.3 where such an extension is implemented for the Incremental Effect Additivity (IEA) synergy theory; the same idea works equally well for the extended Loewe additivity.

Eq. (1.3.2) can be represented in an equivalent form

$$\sum_{j=1}^{N} \frac{d_j}{F_j(E)} = 1, \tag{1.3.5}$$

where $d_1, d_2, ..., d_N$ are the doses of $N$ agents whose combined action in the absence of interaction produces effect $E$. The set of all such dose vectors $(d_1, d_2, ..., d_N)$ for a fixed effect $E$ forms a polygonal region in an $(N-1)$-dimensional hyperplane in $R^N$ determined by Eq. (1.3.5). In the case of two agents, typically considered in the literature, Eq. (1.3.5) describes a segment in the plane called an *isobole* (see e.g. Tallarida 2016). Actual experiments with the joint action of drugs, toxins and other agents without known interactions reveal that in many cases the aforementioned isoeffective regions are non-linear (see e.g. Lederer et al. 2016). This represents a major flaw of Loewe additivity principle, which calls for alternative synergy theories. One of them, Incremental Effect Additivity (IEA) introduced in Ham et al. (2018), is studied in this work.

*1.3.3 Specialized Synergy Theories*

There are other suggested replacements for SEA synergy theory which obey the sham mixture principle and are useful, but are applicable only if all 1-agent DERs in a mixture have a specific functional form – such as LQ functions. Examples include the formalisms described in Zaider and Rossi (1980), Zaider (1990) and Chou (2006). Our paper will not discuss these specialized synergy theories further, focusing instead on a more flexible IEA formalism.

**1.4 IEA Synergy Theory**

IEA synergy theory obeys the sham mixture principle as well as the more general mixture of mixtures principle and can handle readily multi-component mixtures whose 1-agent components' DERs are highly nonlinear and heterogeneous. "Incremental" in IEA refers to the fact that the first derivatives of 1-agent DERs and EDRs play an important role in IEA synergy theory. The underlying idea was suggested by the the radiobiologist G.K. Lam in Lam (1987) – a one-ion DER slope defines a *linear* relation between a sufficiently small dose increment and the corresponding effect increment, thereby to some extent circumventing the nonlinearities that plague SEA synergy theory. The Lam's replacement for SEA proposed in (Lam 1994), however, does not obey the sham mixture principle.

**1.5 Preview**

Section 2 below describes mathematical foundations of the IEA theory and its heuristic justification. Section 3 presents our main results and discusses them. Sections 2 and 3 focus on mathematical properties of synergy theories – properties that (contrary to statistical properties) hold whatever data is being used to model 1-agent DERs. The final section, Discussion and Conclusions, provides a summary and delivers a broader perspective. Programs that generated Figs 1-11 in the form of R and Matlab scripts are given in Supplementary Materials.

### 1.6 Acronyms and Terminology

The following acronyms and terminology emphasize concepts that are central in this paper:

**DER** – Dose-Effect Relation; **EDR** – Effect-Dose Relation (for a monotonic increasing DER, the inverse function which gives dose as a function of effect); **Generator** – the reciprocal of the derivative of an EDR; **IEA** – Incremental Effect Additivity; **IVP** – Initial Value Problem; **NSNA** – Neither Synergy Nor Antagonism; **ODE** – Ordinary Differential Equation; **SEA** – Simple Effect Additivity synergy theory.

## 2 Mathematical Methodology

Recall that synergy theories focus on calculating a NSNA mixture DER from the 1-agent mixture components' DERs and sometimes use EDRs to facilitate those calculations. This section focuses on conceptual, mathematical and computational aspects of this approach.

### 2.1 Basic Assumptions on DERs

Recall that by definition a 1-agent DER $E(d)$ has domain $[0, \infty)$ and obeys $E(0) = 0$. Most synergy theories including Loewe additivity require additional conditions, notably continuity and monotonic increase, on DERs. Because the behavior of any 1-agent DER at very small doses cannot be empirically determined with certainty, in this article we seek to avoid imposing strong assumptions on such behavior including differentiability of DERs at $d = 0$. Henceforth we make the following requirements:

  *(a) Each 1-agent DER function $E_j(d)$ is continuous at $d = 0$ and satisfies $E_j(0) = 0$, $1 \leq j \leq N$;*

  *(b) Every DER $E_j$ is piecewise continuously differentiable on $(0, \infty)$ with no more than finitely many points of discontinuity of the derivative $E'_j$ on any finite interval $(0, A)$;*

  *(c) For each $j$, $E'_j(d)$ is finite and positive for all $d > 0$, where the derivative may be one- or two-sided.*

In what follows we will consider first a simpler case where all 1-agent DERs have the *same* range $[0, h)$, where $h$ is finite or infinite. A more general case of distinct ranges will be addressed in subsection 3.3.

Observe that if $\varphi$ is a $C^1$-diffeomorphism of $[0, h)$ then functions $\varphi \circ E_j$ satisfy conditions (a-c) and can therefore serve as DERs. This allows one to standardize, in certain settings, the range of effect to $[0,1)$ or

[0, ∞). The above invariance property also holds for the functions $E_j \circ \psi$, where $\psi$ is a $C^1$-diffeomorphism of [0, ∞). As indicated in subsection 1.1, in most cases such transformations $\psi$ are limited to dilations. Finally, under assumptions (a-c) DERs $E_j$ are bijective and, importantly, the corresponding EDR functions $F_j$: [0, h) → [0, ∞), where $F_j = E_j^{-1}$, also meet conditions (a-c).

## 2.2 The IEA Equation

In IEA synergy theory we assume that, in the absence of synergy and antagonism between the agents, the NSNA DER function $I(d)$ for the mixture of $N \geq 2$ agents with fractional weights $r_j$, $1 \leq j \leq N$, is the solution of the following autonomous IVP:

$$I' = \sum_{j=1}^{N} r_j E_j'[E_j^{-1}(I)], \quad I(0) = 0. \tag{2.2.1}$$

Using $E_j'[E_j^{-1}(I)]$, with $I$ the effect due to the entire mixture, in Eq. (2.2.1) is the key assumption of IEA synergy theory. In applications to radiobiology, Eq. (2.2.1) can be interpreted heuristically as follows (Ham et al. 2018). As the total mixture dose $d$ increases slightly, every individual component dose $d_j$ has a slight proportional increase since $d_j = r_j d$. Therefore, every mixture component contributes some incremental effect. The size of the incremental effect is determined by the state of the biological target, specifically by the total effect $I(d)$ already contributed by all the components acting jointly – and *not* by the dose (or the effect) the individual component has already contributed. The intuitive picture is the following. When an incremental dose arrives each mixture one-ion component has to "decide" how much incremental effect to contribute. The component cannot use the dose it itself has already contributed, much less the total mixture dose already contributed by all the one-ion components collectively, to decide – those doses have already come and gone. But the biological target does "remember" the total effect that all one-ion components acting together have already contributed, and this systemic property can be used to determine each component's incremental effect – each component can adjust to its own influence and the influence of all the other components.

In the right-hand side of Eq. (2.2.1) effect $I$ is acting as the control variable and $d$ is acting as the response variable instead of vice-versa. Because $I$ is a systemic property and $d$ is not, this switch represents an important conceptual improvement.

## 2.3 Mathematical Theory of the IEA Equation

Eq. (2.2.1) can be applied to a single agent with DER $I = E(d)$, which yields

$$I' = E'[E^{-1}(I)], \quad I(0) = 0. \tag{2.3.1}$$

In this equation, the derivative of the agent's DER is expressed through the effect rather than dose. The function $G(I) = E'[E^{-1}(I)]$ associated with a given agent will be called this agent's *generator*. In terms of generators, the IEA equation, Eq. (2.2.1), takes on the form

$$I' = \sum_{j=1}^{N} r_j G_j(I), \quad I(0) = 0. \tag{2.3.2}$$

Thus, the general form of the IEA equation becomes

$$I' = G(I), \quad I(0) = 0, \tag{2.3.3}$$

where $G = \sum_{j=1}^{N} r_j G_j$ is the generator of the mixture of $N$ agents.

In this subsection, we develop a general mathematical theory of Eq. (2.3.3) focusing on the existence, uniqueness and specific form of its solution. The main result is given by the following theorem, which represents a formal justification for the method of separation of variables in Eq. (2.3.3).

**Theorem 2.3.1.** *Suppose $G: (0, \infty) \to (0, \infty)$ is a continuous function such that*

(a) $\int_0^x \frac{du}{G(u)} < \infty$ *for all* $x > 0$ *and*

(b) $\int_0^\infty \frac{du}{G(u)} = \infty.$

*Then the inverse function to*

$$F(x) = \int_0^x \frac{du}{G(u)}, \quad x > 0, \tag{2.3.4}$$

*represents a unique solution $I$ of the IVP (2.3.3) defined on $[0, \infty)$ such that $I(d) > 0$ for all $d > 0$.*

*Proof.* Set $F(0) = 0$. It follows from conditions (a), (b) of the theorem that function $F$ is strictly increasing, continuous at $x = 0$, continuously differentiable on $(0, \infty)$, and has $[0, \infty)$ as its range. Thus, $F: [0, \infty) \to [0, \infty)$ is a bijection. Denote by $I$ the inverse function for $F$. Then $I(0) = 0$, $I(d) > 0$ for all $d > 0$, and for every $d > 0$ we have in view of the formula for the derivative of the inverse function and Eq. (2.3.4)

$$I'(d) = \frac{1}{F'(I(d))} = G(I(d)).$$

Therefore, function $I$ is a required positive solution of the IVP (2.3.3).

To show uniqueness, suppose $I: [0, \infty) \to [0, \infty)$ is a solution of the IVP (2.3.3) such that $I(d) > 0$ for all $d > 0$. Then $I(0) = 0$ and $I'(u) = G(I(u))$ for all $u > 0$. Equivalently,

$$\frac{I'(u)}{G(I(u))} = 1 \quad \text{for all} \quad u > 0.$$

Integrating this equation over $[0, d]$ and using the Fundamental Theorem of Calculus we find that $F(I(d)) = d$ for every $d \geq 0$. Therefore, $I = F^{-1}$. The proof is completed.

**Remark 1**. Denote $G(0) = \lim_{x \to 0+} G(x)$ assuming the limit exists. If $G(0) = 0$ and function $G$ meets all the assumptions of Theorem 2.3.1 then, along with a positive solution $I$, Eq. (2.3.3) has the trivial solution $I_0 = 0$.

Furthermore, for every $d_0 > 0$, these two solutions can be combined into a new solution, $I_1$, by setting $I_1(d) = 0$ for $0 \leq d \leq d_0$ and $I_1(d) = I(d-d_0)$ for $d > d_0$.

**Remark 2**. One convenient class of generators that satisfy the conditions of Theorem 2.3.1 and produce a unique solution of the IVP (2.3.3) are continuous functions on $(0, \infty)$ that are uniformly bounded above and below by positive constants. A more general class of generators with this property is the set of continuous positive functions $G$ on $[0, \infty)$ such that $G(x) \leq Cx^\beta$ with some constants $C > 0$ and $\beta \leq 1$ for all sufficiently large $x > 0$.

**Remark 3.** Notice that apart from the assumed continuity, Theorem 2.3.1 does not require any additional smoothness of function $G$ including commonly presumed Lipschitz condition or a stronger condition of the existence of uniformly bounded derivative.

**Remark 4.** Theorem 2.3.1 has an immediate extension to the case where for some finite number $h > 0$, generator $G: (0, h) \to (0, \infty)$ is a continuous function such that $\int_0^x \frac{du}{G(u)} < \infty$ for all $x \in (0, h)$ and $\int_0^h \frac{du}{G(u)} = \infty$. In this case, the unique positive solution $I$ of the IVP (2.3.3) takes values in $[0, h)$ rather than $[0, \infty)$. If $G$ is a 1-agent generator then its DER $E$ has the range $[0, h)$.

**Remark 5.** If $A = \int_0^\infty \frac{du}{G(u)} < \infty$ (or $A = \int_0^h \frac{du}{G(u)} < \infty$, see Remark 4) then the positive solution, $I$, of the IVP (2.3.3) provided by Theorem 2.3.1 is only defined on the interval $[0, A)$ and $\lim_{d \to A-} I(d) = \infty$. In the case where $G$ is a 1-agent generator with DER $E$ the interpretation of this possibility is that the finite dose $A$ of the agent brings about a catastrophic damage to the agent's target.

**Remark 6.** Suppose $G: [0, \infty) \to [0, \infty)$ with $G(0) = 0$ is a continuous function such that $G(x) > 0$ for $x > 0$. It follows from the proof of Theorem 2.3.1 that if $\int_0^x \frac{du}{G(u)} = \infty$ for all $x > 0$ then IVP (2.3.3) has only trivial solution.

**Remark 7.** Theorem 2.3.1 and all the above remarks remain true if function $G: (0, \infty) \to (0, \infty)$ is piecewise continuous with only finitely many jump discontinuities on any finite interval $(0, a)$. In this case, equality $I'(d) = G(I(d))$ is assumed to be true for all $d > 0$ such that $I(d)$ is a point of continuity of function $G$.

We now illustrate Theorem 2.3.1 with the following three examples of single-agent DERs.

(1) Suppose $E(d) = d^a$, $a > 0$, then Eq. (2.3.1) becomes $I' = aI^{1-1/a}$, $I(0) = 0$. Clearly, function $G(u) = au^{1-1/a}$ satisfies all the conditions of Theorem 2.3.1. Therefore, $I(d) = d^a$ is the only positive solution of Eq. (2.3.1). However, for $a > 1$ this equation also has the extraneous trivial solution $I = 0$.

(2) Consider the "extremely flat at $d = 0$" DER $E(d) = \exp(-d^{-a})$ with $a > 0$ and range $[0, 1)$ that produces the IVP $I' = aI [\ln(1/I)]^{1+1/a}$, $I(0) = 0$. In this case, the generator function $G(u) = au [\ln(1/u)]^{1+1/a}$ is defined for $0 < u < 1$ and satisfies the conditions of Remark 4 with $h = 1$. Thus, $I = E(d)$ is the only positive solution of the IVP (2.3.1). However, for any $a > 0$ this equation also has the trivial solution $I = 0$.

(3) Let $E(d) = d/(d+1)$ with range $[0, 1)$, then $G(u) = (1-u)^2$, $0 \leq u < 1$. Here again the generator satisfies the conditions of Remark 4 with $h = 1$. However, in this case $I = 0$ is not a solution of the IVP (2.3.1), and so the DER in question is the unique solution of this IVP.

## 2.4 An EDR Formulation and Solution of the IEA Equation

Recall that the inverse function, $F$, for a 1-agent DER, $E$, is called the agent's EDR. By the formula for the derivative of the inverse function we have for $I > 0$

$$F'(I) = \frac{1}{E'[E^{-1}(I)]} = \frac{1}{G(I)}. \tag{2.4.1}$$

Hence by integration and using the property $F(0) = 0$ we find that $F(I) = \int_0^I \frac{du}{G(u)}$, $I > 0$, compare with Eq. (2.3.4). Thus, in the case of an individual agent, the general premises of the IEA theory, see subsection 2.1, can equivalently be expressed in terms of generators as assumptions of Theorem 2.3.1 or its extensions given in Remark 4 (the case of finite h) and/or Remark 7 (the case of piecewise continuous generators).

We now address Eq. (2.2.1) for the mixture DER of $N$ agents, which can be represented in the form of Eq. (2.3.2). Observe that in this case the mixture generator $G$ is continuous (or piecewise continuous). For EDRs of individual agents we have similar to Eq. (2.4.1)

$$F_j'(I) = \frac{1}{E_j'[E_j^{-1}(I)]} = \frac{1}{G_j(I)}, \quad 1 \leq j \leq N. \tag{2.4.2}$$

Eq. (2.4.1) equally applies to the mixture EDR $F$ and the corresponding mixture DER $E$. Therefore, the mixture EDR satisfies the following equation:

$$\frac{1}{F'(I)} = \sum_{j=1}^N \frac{r_j}{F_j'(I)}, I > 0; \quad F(0) = 0. \tag{2.4.3}$$

This is the EDR version of the IEA equation, Eq. (2.2.1). Verification of the mixture of mixtures property for IEA based on Eqs (2.2.1) or (2.4.3) amounts to a fairly trivial exercise in linear algebra.

From Eq. (2.4.3) the mixture EDR can formally be found by integration:

$$F(I) = \int_0^I \left[ \sum_{j=1}^N \frac{r_j}{F_j'(x)} \right]^{-1} dx. \tag{2.4.4}$$

Whether function $F$ actually represents a solution to the IEA equation (2.3.2) depends, therefore, on whether conditions (a) and (b) of Theorem 2.3.1 (or Remark 4) are satisfied.

We begin with condition (a), i.e. that $F(I)$ is finite or all $I > 0$. Note that in the case where $F_j'(0) = \infty$ (or equivalently if $E_j'(0) = 0$) for all $j$ the Riemann integral in Eq. (2.4.4) should be treated as improper.

**Proposition 2.4.1.** *Suppose individual DERs satisfy assumptions (a-c). Then $F(I) < \infty$ for all $I \in (0, h)$.*

*Proof.* For vector $\boldsymbol{x} = (x_1, x_2, ..., x_N)$ with positive components, non-zero number $a$, and normalized positive weight sequence $\boldsymbol{r} = (r_1, r_2, ..., r_N)$ as above, denote by $A_a(\boldsymbol{x}, \boldsymbol{r})$ the weighted average of $\boldsymbol{x}$ of order $a$ defined by

$$A_a(\boldsymbol{x}, \boldsymbol{r}) = \left( \sum_{j=1}^N r_j x_j^a \right)^{1/a}. \tag{2.4.5}$$

Recall that for any fixed vectors $x$ and $r$, $A_a(x, r)$ is an increasing function of $a$ (Hardy et al. 1952). In particular, setting $a = -1$ and $a = 1$ we have $H(x, r) = A_{-1}(x, r) \leq A_1(x, r)$, which represents the inequality between the weighted harmonic and arithmetic means. Applying this inequality to vector $x = (F_1'(I), F_2'(I), \ldots, F_N'(I))$ and using Eq. (2.4.3) and assumptions (b, c) from subsection 2.1 we find that

$$F'(I) = \left[\sum_{j=1}^{N} \frac{r_j}{F_j'(x)}\right]^{-1} \leq \sum_{j=1}^{N} r_j F_j'(I), \quad 0 < I < h. \tag{2.4.6}$$

For any $\varepsilon \in (0, I)$, we integrate this inequality and use assumptions (a-c) to obtain

$$\int_\varepsilon^I \left[\sum_{j=1}^{N} \frac{r_j}{F_j'(x)}\right]^{-1} dx \leq \int_\varepsilon^I \sum_{j=1}^{N} r_j F_j'(x) dx = \sum_{j=1}^{N} r_j [F_j(I) - F_j(\varepsilon)] \leq \sum_{j=1}^{N} r_j F_j(I).$$

Therefore, from Eq. (2.4.4) we infer that

$$F(I) = \lim_{\varepsilon \to 0} \int_\varepsilon^I \left[\sum_{j=1}^{N} \frac{r_j}{F_j'(x)}\right]^{-1} dx \leq \sum_{j=1}^{N} r_j F_j(I) < \infty. \tag{2.4.7}$$

The proof is completed.

We now turn to a fundamental, yet more subtle, question regarding the validity of condition (b) of Theorem 2.3.1, i.e. whether

$$F(\infty) = \int_0^\infty \left[\sum_{j=1}^{N} \frac{r_j}{F_j'(x)}\right]^{-1} dx = \infty. \tag{2.4.8}$$

The following example demonstrates that even under the assumptions (a-c) of the IEA theory this does not necessarily have to be the case.

**Example 1**. Let functions $f, g: [0, \infty) \to (0, 1]$ be defined by $f(x) = 1/(1+x)$ and $g(x) = 1/(1+x)^2$. Partition their domain $[0, \infty)$ into intervals $[n, n+1)$ for integer $n \geq 0$. Let $A_n$ be such an interval for even $n$ and $B_n$ for odd $n$. Let $F_1'(x) = f(x)$ for $x \in A_n$ and $F_1'(x) = g(x)$ for $x \in B_n$; symmetrically, let $F_2'(x) = g(x)$ for $x \in A_n$ and $F_2'(x) = f(x)$ for $x \in B_n$. Thus, the generators of the two agents are switching between $1+x$ and $(1+x)^2$ on intervals $A_n$ and $B_n$. Note that functions $F_1'$ and $F_2'$ are positive and piecewise continuous with finitely many jump discontinuities on any finite interval. Let $F_i(x) = \int_0^x F_i'(t) dt$, $i = 1, 2$. Observe that

$$F_1(\infty) = \sum_{n=0}^{\infty} \int_{2n}^{2n+1} \frac{dx}{1+x} + \sum_{n=0}^{\infty} \int_{2n+1}^{2n+2} \frac{dx}{(1+x)^2} = \sum_{n=0}^{\infty} \ln\left(1 + \frac{1}{2n+1}\right) + \sum_{n=0}^{\infty} \frac{1}{(2n+2)(2n+3)} = \infty$$

because the first series diverges and the second converges. Similarly, $F_2(\infty) = \infty$. Therefore, functions $F_1$ and $F_2$ satisfy conditions (a-c) of the IEA theory with $h = \infty$, and so their inverse functions can serve as legitimate individual DERs for two agents. We now show that the DER of the 1:1 mixture of these agents "blows up" at some finite dose. According to Remark 5 to Theorem 2.3.1 and Eq. (2.4.8) we only have to verify that

$$\int_0^\infty \frac{2}{\frac{1}{F_1'(x)} + \frac{1}{F_2'(x)}} dx < \infty.$$

In fact,

$$\int_0^\infty \frac{2}{\frac{1}{F_1'(x)} + \frac{1}{F_2'(x)}} dx = \int_0^\infty \frac{2}{\frac{1}{f(x)} + \frac{1}{g(x)}} dx = \int_0^\infty \frac{2}{(1+x)(2+x)} dx = 2\ln 2.$$

Therefore, for the mixture DER, $E$, we have $\lim_{d \to 2\ln 2-} E(d) = \infty$.

Example 1 demonstrates that indiscriminate use of individual DERs may occasionally lead to the biologically impossible "blow-up" effect where in the absence of synergy and antagonism as defined by the IEA equation the mixture of several "regular" DERs (i.e. with finite effect for any finite dose) may prove "singular," i.e. produce infinite effect at finite dose. To avoid the "blow-up" phenomenon, individual EDRs must satisfy the condition

$$F(h) = \int_0^h \left[ \sum_{j=1}^N \frac{r_j}{F_j'(x)} \right]^{-1} dx = \infty, \tag{2.4.9}$$

compare with Eq. (2.4.8), or equivalently, when expressed through individual generators,

$$\int_0^h \frac{dx}{\sum_{j=1}^N r_j G_j(x)} = \infty.$$

Interestingly, the "blow-up" condition (2.4.9) can be stated in an equivalent form that only contains individual EDRs but is free of their weights:

**Proposition 2.4.2.** *Condition (2.4.9) is equivalent to*

$$\int_0^h \min_{1 \le j \le N} F_j'(x)\, dx = \infty. \tag{2.4.10}$$

*Proof.* Consider the following estimates for the mixture generator $G(x) = \sum_{j=1}^N r_j G_j(x)$:

$$r(x) \max_{1 \le j \le N} G_j(x) \le \sum_{j=1}^N r_j G_j(x) \le \max_{1 \le j \le N} G_j(x), \quad 0 < x < h, \tag{2.4.11}$$

where $r(x)$ is the sum of weights $r_i$ over indices $i$ such that $G_i(x) = \max_{1 \le j \le N} G_j(x)$. Using Eq. (2.4.3) for the mixture EDR, Eq. (2.4.2) and the estimate $r(x) \ge \min_{1 \le j \le N} r_j = \rho > 0$ we represent inequalities (2.4.11) through EDRs in the form

$$\min_{1 \le j \le N} F_j'(x) \le F'(x) \le \rho^{-1} \min_{1 \le j \le N} F_j'(x), \quad 0 < x < h,$$

whence by integration we find that

$$\int_0^I \min_{1 \le j \le N} F_j'(x)\, dx \le F(I) \le \rho^{-1} \int_0^I \min_{1 \le j \le N} F_j'(x)\, dx, \quad 0 < I < h. \tag{2.4.12}$$

In particular,

$$\int_0^h \min_{1 \le j \le N} F_j'(x)\, dx \le F(h) \le \rho^{-1} \int_0^h \min_{1 \le j \le N} F_j'(x)\, dx.$$

Therefore, $F(h) = \infty$ iff condition (2.4.10) is met, which completes the proof.

Equivalently, in terms of generators Eq. (2.4.10) takes on the form

$$\int_0^h \frac{dx}{\max_{1\le j\le N} G_j(x)} = \infty.$$

We conclude this subsection with an example illustrating Theorem 2.3.1.

**Example 2**. Consider a 1:1 mixture ($r_1 = r_2 = 0.5$) of two agents with DERs $E_1(d) = d^2$ and $E_2(d) = d^3$, where $h = \infty$. In this case $F_1(I) = I^{1/2}$, $F_2(I) = I^{1/3}$, $G_1(I) = 2I^{1/2}$, $G_2(I) = 3I^{2/3}$ and from Eq. (2.4.4) we obtain for $I \ge 0$

$$F(I) = 2\int_0^I \frac{dx}{2x^{1/2} + 3x^{2/3}}.$$

Clearly, the singularity of the integrand at $x = 0$ is integrable and the "no blow-up" condition (2.4.8) is satisfied. By a rationalizing substitution $x = t^6$ we find that

$$F(I) = 12\int_0^{I^{1/6}} \frac{t^2 dt}{3t + 2} = 2I^{\frac{1}{3}} - \frac{8}{3}I^{\frac{1}{6}} + \frac{16}{9}\ln\left(1 + 1.5I^{\frac{1}{6}}\right). \tag{2.4.13}$$

The 1-agent EDRs and the mixture EDR $F(I)$ are shown graphically in Fig. 2.

## 2.5. Comparison of DERs and EDRs

All properties of DERs can be expressed in terms of EDRs and vice versa. Here are some examples:

(A) Let $E_1$ and $E_2$ with $E_1(0) = E_2(0) = 0$ be two continuous increasing DER functions having the same range $[0, h]$. Then $E_1(d) \le E_2(d)$ for all $d \ge 0$ if and only if $F_1(I) \ge F_2(I)$ for all $I \in [0, h]$;

(B) The graphs of two continuous increasing DER curves $E_1(d)$ and $E_2(d)$ intersect at $d = d_0 > 0$ if and only if the respective EDR curves $F_1(I)$ and $F_2(I)$ intersect at $I = I_0$, where $I_0 = E_1(d_0) = E_2(d_0)$;

(C) An increasing DER is convex/concave if and only if the corresponding EDR is concave/convex. This follows from the fact that the graph of a DER function $I = E(d)$ is at the same time the graph of the corresponding EDR function $d = F(I)$ if we view the $E$ axis as the axis for independent variable. Thus, a chord generated by two distinct points on the graph of the DER function stays above/below the graph of the DER function if and only if the same chord stays below/above the graph of the EDR function.

In spite of theoretical equivalence between DERs and EDRs, the mixture EDR function $F$ given by Eq. (2.4.4) has a number of practical advantages over the mixture DER function $E$ defined as a solution of the IVP (2.2.1):

(1) The derivatives of all 1-agent EDR functions in Eq. (2.4.3) are evaluated at the same effect $I$ while in Eq. (2.2.1) the derivatives of individual DERs are evaluated at different doses. This facilitates establishing various properties of NSNA mixture DERs and EDRs. One of them is the invariance of the NSNA criterion (2.4.3) under $C^1$-diffeomorphisms of the effect range $[0, h]$. In fact, if $\varphi$ is such a diffeomorphism then a routine application of the Chain Rule would show that if EDR functions $F_j$, $1 \le j \le N$, and $F$ satisfy Eq. (2.4.3) then so do functions $F_j \circ \varphi$, $1 \le j \le N$, and $F \circ \varphi$. Establishing an equivalent invariance property for Eq. (2.2.1) requires more work.

(2) The mixture EDR can directly be found from Eq. (2.4.3) by integration, see Eq. (2.4.4). As shown in subsections 2.3 and 2.4, this circumvents, under conditions (a-c) and assumption (2.4.9), subtle questions about the existence, uniqueness and regularity of the solution to the IVP (2.2.1).

(3) In many cases the mixture EDR can, while the mixture DER cannot, be computed in closed form, see e.g. Example 2. Other examples illustrating the same phenomenon will come to light in what follows. The advantage of closed forms is that they give insights into analytic and statistical properties of mixture EDR as a function defined on the entire parameter space, a job for which purely numerical methods are ill-suited. That is why most of the figures illustrating our results in section 3 display EDR plots.

## 2.6 Preview Example: Solving the NSNA Equations for Three Synergy Theories

As a summary of the EDR/DER comparison in subsection 2.6, and as a preview of the examples discussed in section 3, Fig. 3 presents an example of a 4-component mixture where each component agent contributes equally to the total mixture dose $d$ and has a smooth, monotonic increasing DER. Two of the 1-agent DERs are concave, two are convex. Fig. 3 features NSNA curves obtained using three different synergy theories (SEA, Lowe additivity and IEA).

## 3. Mathematical Properties of IEA Synergy Theory: Results, Examples and Counterexamples

In subsections 3.1 and 3.2 we assume that all 1-agent DERs have the same range. Subsection 3.3 deals with a more general case where some of the ranges may be distinct and with complications that then arise.

### 3.1 Generalized Similarity Theory

We illustrate the theory developed in subsection 2.4 with the case where individual DERs are obtained by scaling from the same DER, i.e.

$$E_j(d) = P_j \varphi(d), \ 1 \leq j \leq N, \tag{3.1.1}$$

where $P_j$ are positive coefficients and function $\varphi: [0, \infty) \to [0, h)$ has properties (a)-(c) postulated in subsection 2.1. Let $\psi: [0, h) \to [0, \infty)$ be the inverse function for $\varphi$. Then for the 1-agent EDRs we have

$$F_j(I) = \psi\left(\frac{I}{P_j}\right), \ 0 \leq I < h.$$

In view of Eq. (2.4.4) the EDR for the combined action of $N$ agents under the NSNA condition is

$$F(I) = \int_0^I \left[\sum_{j=1}^N \frac{r_j P_j}{\psi'\left(\frac{x}{P_j}\right)}\right]^{-1} dx. \tag{3.1.2}$$

Discussed below are two cases where this integral can be computed in closed form.

**Case 1.** $\varphi(d) = d^a$, where $a > 0$.

In this case $\psi(I) = I^{1/a}$, and from Eq. (3.1.2) we find that

$$F(I) = \int_0^I \left[\sum_{j=1}^N \frac{ar_j P_j}{\left(\frac{x}{P_j}\right)^{\frac{1}{a}-1}}\right]^{-1} dx = \bar{P}_{1/a}^{-1/a} \int_0^I a^{-1} x^{1/a-1} dx = \bar{P}_{1/a}^{-1/a} I^{1/a},$$

where

$$\bar{P}_{1/a} = \left(\sum_{j=1}^N r_j P_j^{1/a}\right)^a$$

is the weighted average of order 1/a of the vector of coefficients ($P_1$, $P_2$, …, $P_N$), compare with Eq. (2.4.5). Therefore, the mixture DER is given by the formula

$$E(d) = \bar{P}_{1/a} d^a, \tag{3.1.3}$$

i.e. is of the same kind as individual DERs, see Eq. (3.1.1). Clearly, the mixture EDR is sandwiched between respective individual EDRs with the smallest and largest coefficients $P_j$. Eq. (3.1.3) with $a = 1$ represents the main result of the classic linear similarity theory (Berenbaum, 1989).

**Case 2.** $\varphi(d) = \ln(Ad + 1)$, $A > 0$.

Then $\psi(I) = A^{-1}[\exp(I)-1]$, and in view of Eq. (3.1.2)

$$F(I) = A^{-1} \int_0^I \left(\sum_{j=1}^N r_j P_j e^{-x/P_j}\right)^{-1} dx.$$

This integral can be computed in closed form only in a few particular instances. One of them occurs when $N = 2$ and one of the coefficients $P_1$, $P_2$ is twice as large as the other. Suppose $P_2 = P$ and $P_1 = 2P$, then setting $r = r_1$ and making a change of variable $u = \exp\{x/2P\}$ we have

$$F(I) = \frac{1}{A}\int_0^I \left[2Pre^{-x/2P} + P(1-r)e^{-x/P}\right]^{-1} dx = \frac{1}{AP}\int_0^I \frac{e^{x/P}dx}{2re^{x/2P}+1-r}$$

$$= \frac{2}{A}\int_1^{e^{I/2P}} \frac{u\,du}{2ru+1-r} = \frac{1}{Ar}\left[e^{I/2P} - 1 - \frac{1-r}{2r}\ln\left(1 + \frac{2r}{1+r}(e^{I/2P} - 1)\right)\right].$$

In this case, yet again, the mixture EDR is, while the mixture DER is not, computable in closed form.

In particular, Case 2 leads to the following specific example:

**Example 3.** Set $N = 2$, $r = 0.5$ and $A = P = 1$, then

$$F(I) = 2(e^{I/2} - 1) - \ln\left(1 + \frac{2}{3}(e^{I/2} - 1)\right). \tag{3.1.4}$$

The graphs of the individual EDRs $F_1(I) = \exp(I/2)-1$, $F_2(I) = \exp(I)-1$ and the mixture EDR $F(I)$ are shown in Fig. 4.

### 3.2 The Betweenness Property for Mixture IEA EDRs

In this section, we study the betweenness property and discover a surprising asymmetry between the upper and lower bounds for the mixture EDR. Because betweenness violation by the NSNA IEA DER curve is a newly discovered, important flaw of IEA synergy theory, we set out to delineate specifically when violation occurs and what conditions on 1-agent DERs insure betweenness. Assumptions (a)-(c) of subsection 2.1 are a general premise for all the results in this subsection.

**Proposition 3.2.1.** *The N-agent mixture EDR F(I) satisfies the inequality*

$$F(I) \leq \max\{F_j(I): 1 \leq j \leq N\}, \quad 0 \leq I < h. \tag{3.2.1}$$

*Proof.* This inequality immediately follows from inequality (2.4.7).

Thus, under the IEA condition, the mixture EDR curve cannot cross into the region above the upper envelope of individual EDR curves. Equivalently,

$$E(d) \geq \min\{E_j(d): 1 \leq j \leq N\}, \quad d \geq 0.$$

A natural question is whether an inequality dual to (3.2.1) holds for the lower envelope of individual EDR curves. Example 2 in subsection 2.5 gives a negative answer to this question, see Fig. 2. In fact, the two individual EDR curves in Example 2 intersect at point (1, 1) while for the mixture EDR we have

$$F(1) = \frac{16}{9} \ln 2.5 - \frac{2}{3} \cong 0.962 < 1.$$

In Example 2, derivatives of *both* individual DERs at $d = 0$ vanished. To make sure this is inessential for violation of the betweenness property, consider the case where $N = 2$, $h = \infty$, $E_1(d) = ad$ and $E_2(d) = bd^2$ with $a, b > 0$. Then $F_1(I) = I/a$ and $F_2(I) = \sqrt{I/b}$. Applying Eq. (2.4.3) and setting $r_1 = r$, we obtain

$$F'(I) = \frac{1}{ar(1+C\sqrt{I})}, \quad I \geq 0,$$

where $C = 2(1-r)\sqrt{b}/(ar)$. Then

$$F(I) = \frac{1}{ar}\int_0^I \frac{dx}{1+C\sqrt{x}} = \frac{2}{arC^2}\int_1^{1+C\sqrt{I}} \frac{(u-1)du}{u} = \frac{2}{arC^2}[C\sqrt{I} - \ln(1 + C\sqrt{I})],$$

where we used a substitution $u = 1 + C\sqrt{x}$. Selection of some particular values of parameters $a$, $b$ and $r$ leads to the following example.

**Example 4.** Let $r = 0.5$, $a = b = 1$. Then $C = 2$ and

$$F(I) = 2\sqrt{I} - \ln(1 + 2\sqrt{I}).$$

Note that the lower envelope of the two individual EDRs must pass through the point (1, 1) of their intersection. However, $F(1) = 2 - \ln 3 \cong 0.901 < 1$. The graphs of the functions $F_1$, $F_2$ and $F$ are shown in Fig. 5.

Examples 2 and 4 demonstrate that the betweenness property characteristic of the similarity theory, see subsection 3.1, fails when *distinct* power functions are used for individual DERs. Examination of these examples leads to the following questions, which, to avoid duplication, we pose for EDRs:

(1) Under which conditions does the mixture EDR stay above the lower envelope of individual EDRs?

(2) Is it true that in the case $h = \infty$ the mixture EDR curve eventually (i.e. for sufficiently large effects) enters the region between the lower and upper envelopes of individual EDRs?

(3) When does the mixture EDR curve for two agents pass through the intersection point of their individual EDRs?

The following statement provides a simple sufficient condition for the affirmative answer to question 1 in the case of two agents.

**Proposition 3.2.2.** *Suppose $N = 2$ and $F_1'(I) \leq F_2'(I)$ for $0 < I < h$. Then for the mixture EDR $F(I)$ we have*

$$F_1(I) \leq F(I) \leq F_2(I) \text{ for all } I \in [0, h). \qquad (3.2.2)$$

*Proof.* It follows from Eq. (2.4.3) and inequalities (1.3.4) that under the assumptions of Proposition 3.2.2 $F_1'(I) \leq F'(I) \leq F_2'(I)$ for $0 < I < h$, which after integration yields Eq. (3.2.2).

According to Eq. (2.4.2), universal (i.e. valid for all effects $I \in (0, h)$) ordering of the derivatives of EDRs assumed in Proposition 3.2.2 is equivalent to universal reverse ordering of the generators. Thus, universal ordering of agent generators entails the betweenness property.

An immediate generalization of the betweenness property for mixture EDR expressed by Eq. (3.2.2) to any number of agents is as follows:

**Proposition 3.2.3.** *If for some $k$, $1 \leq k \leq N$, the derivatives of EDRs $F_j$, $1 \leq j \leq N$, $j \neq k$, satisfy the inequality $F_j'(I) \geq F_k'(I)$ for all $I \in [0, h)$ then the mixture EDR has the betweenness property.*

*Proof.* From the assumed inequality for the derivatives of EDRs we conclude by integration that $F_j \geq F_k$ on $[0, h)$ for all $j \neq k$. Therefore, $min\{F_j: 1 \leq j \leq N\} = F_k$. Using Eq. (2.4.3) we find that

$$\frac{1}{F'(I)} = \sum_{j=1}^{N} \frac{r_j}{F_j'(I)} \leq \frac{1}{F_k'(I)}, \quad 0 < I < h.$$

Therefore, $F' \geq F_k'$ on $(0, h)$, which implies through integration that $F \geq F_k = min\{F_j: 1 \leq j \leq N\}$. Combining this with Eq. (3.2.1) establishes the required betweenness property.

**Remark 1.** Under the assumptions of Proposition 3.2.3, $k$-th agent has a universally largest DER and generator, i.e. is a *dominant* agent. Thus, the presence of a dominant agent entails betweenness.

**Remark 2.** A slight modification of the proof of Proposition 2.4.2 would show that if $k$-th agent is dominant then

$$F_k(I) \leq F(I) \leq r_k^{-1} F_k(I), \quad 0 \leq I < h,$$

compare with and Eq. (2.4.12). Thus, a dominant agent imposes robust lower and upper bounds on the mixture EDR that are independent of the number, weights and EDRs of other agents as long as the $k$-th agent remains dominant and retains its weight $r_k$.

Proposition 3.2.2 states that a universal ordering of the derivatives of individual EDRs implies the betweenness property. Is the same true under a weaker condition of universal ordering of individual EDRs? A negative answer to this question, as well as to the above question 2 about eventual betweenness, is provided by the following example.

**Example 5.** Let $N = 2$, $r_1 = r_2 = 0.5$, $h = \infty$, $F_1(I) = I$ and $F_2(I)$ is a piecewise linear function such that $F_2'(I) = 0.2$ for $0 \leq I < 1$, $F_2'(I) = 5$ for $1 \leq I < 1.1$ and $F_2'(I) = 1$ for $I \geq 1.1$. Then

$$F_2(I) = \begin{cases} I/5 & for\ 0 \leq I < 1 \\ 5I - 4.8 & for\ 1 \leq I < 1.1 \\ I - 0.4 & for\ I \geq 1.1 \end{cases} \qquad (3.2.3)$$

A simple computation based on Eq. (2.4.3) yields

$$F(I) = \begin{cases} I/3 & for\ 0 \leq I < 1 \\ 5I/3 - 4/3 & for\ 1 \leq I < 1.1 \\ I - 0.6 & for\ I \geq 1.1 \end{cases}$$

In this case, see Fig. 6, $F_2(I) < F_1(I)$ for all $I > 0$ and the mixture NSNA EDR curve stays below *both* 1-agent EDRs for $I > 1.04$.

We show now that, barring an unlikely special case, the mixture EDR always passes below any intersection point of two individual EDRs, as illustrated by Figs 2 and 5.

**Proposition 3.2.4.** *Suppose $N = 2$ and $A := F_1(I^*) = F_2(I^*)$ for some $I^* > 0$. If the mixture EDR $F(I)$ has the property that $F(I^*) = A$ then $F_1(I) = F_2(I)$ for all $I \in [0, I^*]$.*

*Proof.* Using Eq. (2.4.6), which is based on the inequality between weighted harmonic and arithmetic means, we find that

$$A = F(I^*) = \int_0^{I^*} F'(x)dx \leq \int_0^{I^*} [r_1 F_1'(x) + r_2 F_2'(x)]dx = r_1 F_1(I^*) + r_1 F_2(I^*) = A. \qquad (3.2.4)$$

Thus, the inequality in Eq. (3.2.4) turns into equality. This implies, due to the piecewise continuity of functions $F_1'$ and $F_2'$, that the inequality in Eq. (2.4.6) is actually an equality for all $x \in (0, I^*]$, except possibly for finitely many points. Recall that the inequality between weighted means of two different orders of a vector with positive components turns into equality if and only if the components of the vector are identical (Hardy et al. 1952). Therefore, $F_1'(x) = F_2'(x)$ for all $x \in (0, I^*]$ except possibly for finitely many points. Then by integration we find that $F_1(I) = F_2(I)$ for all $I \in [0, I^*]$, as required.

In spite of Example 5, the universal ordering of DERs within certain particular classes of functions may still imply the betweenness property. One of them is the class of linear-quadratic (LQ) DERs. Recall that LQ DER has the form $E(d) = ad + bd^2$, where $a, b \geq 0$ and $a + b > 0$. For the corresponding EDR and its derivative we have

$$F(I) = \frac{2I}{a + \sqrt{a^2 + 4bI}} \text{ and } F'(I) = \frac{1}{\sqrt{a^2 + 4bI}}.$$

Therefore, if $E_i(d) = a_i d + b_i d^2$, $i = 1, 2$, are two distinct LQ DER such that $a_1 \leq a_2$ and $b_1 \leq b_2$ then the derivatives of the corresponding EDRs are universally ordered, which implies, according to Proposition 3.2.2, the betweenness property. This is illustrated by the following specific example.

**Example 6a.** Suppose $N = 2$, $r_1 = r_2 = 0.5$, $a_1 = 1$, $a_2 = 2$, $b_1 = 0.5$ and $b_2 = 2$. Then the mixture EDR curve lies between the two individual EDRs, see Fig. 7a.

If, on the contrary, $a_1 < a_2$ and $b_1 > b_2$ then the two 1-agent DER curves intersect at $d = (a_2 - a_1)/(b_1 - b_2)$. In this case, Proposition 3.2.4 implies that the betweenness property cannot hold, as illustrated by the following example.

**Example 6b**. Let $N = 2$, $r_1 = r_2 = 0.5$, $a_1 = 1$, $a_2 = 5$, $b_1 = 4$ and $b_2 = 1$. In this case the mixture EDR curve passes below the point of intersection of the two individual EDRs, see Fig. 7b.

Although the mixture EDR for any two LQ DERs can be computed in closed form, the formula is fairly cumbersome; that is why the mixture EDR curves in Fig. 7 were generated using numerical integration.

### 3.3 IEA Synergy Theory for Agents with Distinct Ranges of Effect

Our aim in this subsection is to generalize the above-developed IEA theory to the case where the ranges of effect of the $N \geq 2$ agents are not all identical. Suppose $h_1 < h_2 < \ldots < h_M$, where $2 \leq M \leq N$ and $h_M$ can be finite or infinite, are the agents' distinct maximum effects. Consider all the agents with the range of effect $[0, h_1)$. Renumbering these agents, we may assume these to be agents 1 through $n_1$, where $n_1 \geq 1$. Then $\lim_{d \to \infty} E_j(d) = h_1 < \infty$ for $1 \leq j \leq n_1$. This means that even arbitrarily large doses of any such agent, e.g. agent 1, cannot produce effects $\geq h_1$. Thus, the first $n_1$ terms in the basic IEA Eq. (2.2.1) must vanish for $I \geq h_1$ or, in other words, for such combined effects $I$ the first $n_1$ terms must be dropped from Eq. (2.2.1). Similarly, for $I \geq h_2$ all the additional terms in Eq. (2.2.1) corresponding to the agents whose range of effect is $[0, h_2)$ must be dropped from Eq. (2.2.1). By renumbering these agents, we may assume that these are agents $n_1+1$ through $n_2$, where $n_1 < n_2 \leq N$, etc. Eventually, only the agents with the range of effect $[0, \infty)$, if any, will remain in the IEA equation. A parallel drop-off of terms will occur in Eq. (2.4.3). The same reduction methodology can be applied to Eq. (1.3.2) that describes NSNA mixture EDR under Loewe additivity synergy theory.

To make the described equation reduction compatible with assumptions (b) and (c) from subsection 2.1, we need the following additional assumption:

*(d) If j-th agent, $1 \leq j \leq N$, has finite range of effect then $\lim_{d \to \infty} E'_j(d) = 0$.*

Under this assumption, the drop-off of terms in the RHS of Eq. (2.2.1) does not violate its continuity as a function of $I$.

Condition (d) is satisfied if $E'_j(d)$ is a decreasing function for all, or all sufficiently large, $d$. In particular, this is true for DERs of the form $hd^a/(d^a+b)$ with $a, b > 0$ and $h\exp\{-d^a\}$ with $a > 0$. On the other hand, it is easy to construct a positive continuous integrable function $e$ on $[0, \infty)$ that does not vanish at $\infty$. Then the function $E(d) = \int_0^d e(x)dx$ would meet conditions (a-c) but fail to satisfy condition (d).

*3.3.1 Examples*

Our first example in this subsection deals with a 1:1 mixture of two agents, one with a finite range of effect, which by scaling can be assumed to be [0, 1), and the other with infinite range of effect. Suppose that $E_1(d) = d$ and $E_2(d) = d/(d+a^2)$ with $a > 0$. Note that the two DER curves intersect at some $d \in (0, 1)$ iff $0 < a < 1$. For the corresponding EDRs we have $F_1(I) = I$ and $F_2(I) = a^2I/(1-I)$, so that $F_2'(I) = 1$ and $F_2'(I) = a^2/(1-I)^2$. From Eq. (2.4.3) we find that

$$F'(I) = \begin{cases} \frac{2a^2}{(I-1)^2+a^2} & for\ 0 \leq I < 1 \\ 2 & for\ I \geq 1 \end{cases}$$

whence by integration we obtain

$$F(I) = \begin{cases} 2a\left(\tan^{-1}\frac{1}{a} - \tan^{-1}\frac{1-I}{a}\right) & for\ 0 \leq I < 1 \\ 2\left(I + a\tan^{-1}\frac{1}{a} - 1\right) & for\ I \geq 1 \end{cases}$$

The values $a = \sqrt{3}$ and $a = 1/\sqrt{3}$ produce the following specific examples:

**Example 7a.** Let $a = \sqrt{3}$, in which case individual EDR and DER curves do not intersect. Here

$$F(I) = \begin{cases} \frac{\pi}{\sqrt{3}} - 2\sqrt{3}\tan^{-1}\frac{1-I}{\sqrt{3}} & for\ 0 \leq I < 1 \\ 2\left(I + \frac{\pi}{2\sqrt{3}} - 1\right) & for\ I \geq 1 \end{cases} \qquad (3.3.1a)$$

and, in spite of two different ranges of effect, the mixture EDR curve lies between the two 1-agent EDR curves on the interval [0, 1), see Fig. 8a.

**Example 7b.** Let $a = 1/\sqrt{3}$, in which case individual EDR and DER curves intersect at (2/3, 2/3). Here

$$F(I) = \begin{cases} \frac{2}{\sqrt{3}}\left(\frac{\pi}{3} - \tan^{-1}[\sqrt{3}(1-I)]\right) & for\ 0 \leq I < 1 \\ 2\left(I + \frac{\pi}{3\sqrt{3}} - 1\right) & for\ I \geq 1 \end{cases} \qquad (3.3.1b)$$

Because

$$F(2/3) = \frac{\pi}{3\sqrt{3}} \cong 0.605 < 2/3$$

the mixture EDR curve passes below the point of intersection of individual EDR curves thus violating the betweenness property, as shown in Fig. 8b.

Our next example involves a 1:1 mixture of two agents whose DERs have distinct finite ranges of effect. Let $E_1(d) = d/(d+a)$ and $E_2(d) = 2d/(d+b)$ with $a, b > 0$ be two fractional linear DERs. The two DER curves intersect at some $d > 0$ iff $b > 2a$, in which case $d = b-2a$. For the corresponding EDRs we have $F_1(I) = aI/(1-I)$ and $F_2(I) = bI/(2-I)$. Using Eq. (2.4.3) we find after some algebra that

$$F'(I) = \begin{cases} \frac{p}{(I-q)^2+r^2} & for\ 0 \leq I < 1 \\ \frac{4b}{(2-I)^2} & for\ 1 \leq I < 2 \end{cases}$$

where

$$p = \frac{4ab}{a+2b}, \quad q = \frac{2(a+b)}{a+2b}, \quad r = \frac{\sqrt{2ab}}{a+2b}.$$

Then by integration we obtain

$$F(I) = \begin{cases} 2\sqrt{2ab}\left[\tan^{-1}\frac{\sqrt{2}(a+b)}{\sqrt{ab}} - \tan^{-1}\frac{2(a+b)-(a+2b)I}{\sqrt{2ab}}\right] & \text{for } 0 \le I < 1 \\ 2\sqrt{2ab}\left[\tan^{-1}\frac{\sqrt{2}(a+b)}{\sqrt{ab}} - \tan^{-1}\sqrt{\frac{a}{2b}}\right] + 4b\frac{I-1}{2-I} & \text{for } 1 \le I < 2 \end{cases} \quad (3.3.2)$$

Specialization of these formulas leads to the following two examples.

**Example 8a.** $a = b = 1$. Here individual EDR curves do not intersect and the mixture EDR curve stays between them on the interval [0, 1), see Fig. 9a.

**Example 8b.** $a = 1$, $b = 12$. In this case individual EDR curves intersect at (10/11, 10) and the betweenness property is violated, see Fig. 9b.

Examination of Fig. 7b leads to the following question. Observe that for $I \ge 1$ (i.e. after the term in Eq. (2.4.3) associated with agent 1 has been dropped) the mixture EDR curve stays above the EDR curve for agent 2. Is this always true? In the case $N = 2$ if agent 1 is dropped from the IEA equation for $h_1 \le I < h_2$, the IEA equation on this interval of effects takes on the form $F'(I) = F_2'(I)/r_2$, which implies "eventual betweenness", i.e. that the mixture EDR curve will stay above the EDR curve for agent 2 for all sufficiently large $I$ if $h_2 = \infty$ or for $I$ sufficiently close to $h_2$ if the latter is finite. That the mixture EDR curve doesn't have to always (i.e. not only "eventually") stay above the EDR curve for the remaining agent in the case of finite $h_2$ is demonstrated by Example 8b. In fact, it follows from Eq. (3.3.2) that the mixture EDR and the EDR for agent 2 intersect for $I^* = 2(12-C)/(18-C)$, where

$$C = \sqrt{2ab}\left[\tan^{-1}\frac{\sqrt{2}(a+b)}{\sqrt{ab}} - \tan^{-1}\sqrt{\frac{a}{2b}}\right],$$

which for $a = 1$, $b = 12$ yields $I^* \cong 1.017 > 1$. In the case $h_2 = \infty$ the same effect is illustrated by the following example that modifies Example 5.

**Example 9.** Let $N = 2$, $r_1 = r_2 = 0.5$, $h_1 = 11/9 \cong 1.222$ and $h_2 = \infty$. Suppose $F_2(I)$ is given by Eq. (3.2.3) and

$$F_1(I) = \begin{cases} I & \text{for } 0 \le I < 1.1 \\ \frac{1.21}{11-9I} & \text{for } 1.1 \le I < \frac{11}{9} \end{cases} \quad (3.3.3)$$

Using Eq. (2.4.4) we obtain

$$F(I) = \begin{cases} I/3 & \text{for } 0 \le I < 1 \\ 5I/3 - 4/3 & \text{for } 1 \le I < 1.1 \\ \frac{1}{2} + \frac{11}{15}\left[\tan^{-1}\frac{1}{3} - \tan^{-1}\left(\frac{10}{3} - \frac{30}{11}I\right)\right] & \text{for } 1.1 \le I < \frac{11}{9} \\ 2I + \frac{11}{15}\tan^{-1}\frac{1}{3} - \frac{35}{18} & \text{for } I \ge \frac{11}{9} \end{cases} \quad (3.3.4)$$

For $I = h_1 = 11/9$, where agent 2 is dropped off from the IEA equation (2.4.3), we have

$$F\left(\frac{11}{9}\right) = \frac{1}{2} + \frac{11}{15}\tan^{-1}\frac{1}{3} \cong 0.736$$

while

$$F_2\left(\frac{11}{9}\right) = \frac{11}{9} - 0.4 \cong 0.822.$$

Also, it follows from Eqs (3.2.3) and (3.3.4) that the unique solution of the equation $F(I) = F_2(I)$ on the interval $[11/9, \infty)$ is

$$I^* = \frac{139}{90} - \frac{11}{15}\tan^{-1}\frac{1}{3} \cong 1.308.$$

Therefore, on the interval $[h_1, I^*)$ the mixture EDR curve stays below the remaining EDR curve for agent 1, see Fig. 10.

Finally, to demonstrate the efficiency of the EDR computational methodology developed in this work, we find the combined EDR for a mixture of four agents including one whose DER is "extremely flat" at the origin and has finite range of effect and another with infinite derivative at the origin. Specifically, suppose $E_1(d) = \exp(-1/d)$ for $d \neq 0$ and $= 0$ for $d = 0$, $E_2(d) = \ln(d+1)$, $E_3(d) = d+d^2/2$ and $E_4(d) = d^{1/2}$. Then for the respective 1-agent EDRs we obtain $F_1(I) = 1/\ln(1/I)$, $F_2(I) = \exp(I)-1$, $F_3(I) = (2I+1)^{1/2}-1$ and $F_4(I) = I^2$. Assuming that the doses contributed by the four agents are equal we find from Eq. (2.4.4) that

$$F(I) = 8\int_0^I \frac{xdx}{2(x\ln x)^2 + 2xe^{-x} + 2x\sqrt{2x+1}+1}, \quad 0 \leq I \leq 1. \quad (3.3.5a)$$

Clearly, the integrand doesn't have a singularity at 0 or 1. Furthermore, for $I > 1$

$$F(I) = F(1) + 8\int_1^I \frac{xdx}{2xe^{-x} + 2x\sqrt{2x+1}+1}. \quad (3.3.5b)$$

The 1-agent EDRs and the mixture EDR obtained by numerical integration are shown in Fig. 11.

## 4 Discussion and Conclusions

In this article, we discussed major synergy theories that are widely used in radiation biology, toxicology, pharmacology and other biomedical sciences. The principal focus of this work was rigorous formulation and mathematical analysis of the most recent among these theories, the IEA synergy theory.

### 4.1 SEA and Its Replacements

SEA synergy theory needs to be replaced – synergy theories that fail to obey the sham mixture principle and a more general mixture of mixtures principle are not self-consistent and should be avoided. But our results showed that no fully satisfactory replacement is currently known. The two most promising candidates, Loewe additivity and IEA, are both versatile but have the following cons and pros of their own:

(1) Loewe additivity often fails to provide a good approximation to empirical data on joint effects of non-interacting agents. Also, it cannot deal with mixtures of components whose DERs are not monotonic increasing. On the positive side, we showed the theory always meets the important NSNA mixture DER betweenness criterion.

(2) Like Loewe additivity theory, the basic IEA theory discussed in this article requires monotonicity of the individual DERs. However, IEA often flunks the betweenness test, as our section 3 documents in detail. Furthermore, for some collections of 1-agent regular DERs (i.e. those with finite effect for any finite dose), their mixture, even in the absence of synergy or antagonism as defined by the IEA equation, may produce infinite effect for finite dose, which represents an extreme case of betweenness violation. Such forbidden collections of agents are fully characterized, in terms of their EDRs and generators, in subsection 2.4.

(3) Both synergy theories can handle readily individual DERs with distinct ranges of effect.

## 4.2 IEA Betweenness Violation

In modeling accelerator experiments with mixtures designed to simulate the interplanetary galactic cosmic ray environment, IEA happened to obey betweenness in each considered instance (Huang et al. 2020). The possibility of betweenness violation, discovered and studied in considerable detail in this article, has come as a surprise. Some of the results of this work provide welcome sufficient conditions for betweenness to hold; one of them is Proposition 3.2.3 on dominant agents. Nevertheless, betweenness will henceforth be an issue in applications of IEA.

## 4.3 Future Prospects

High throughput testing of combinations of various substances has produced a copious amount of data, see e.g. Lederer et al. (2019) and Azasi et al. (2020). Biologists and modelers studying such data will surely continue to look for models and theories allowing to detect possible synergy or antagonism. Among the reasons for their importance are the extra dangers brought about by drastic deleterious synergy and the use of synergy theory to plan experiments and interpret their results.

However, it should not be taken for granted that a mixture result confirming expectations from one-agent experiments must necessarily be less valuable than a result contradicting such expectations. For example, stochastic track structure considerations, fundamental in radiobiology, arguably suggest that mixtures of many different densely ionizing radiations may have stochastic ionization patterns not so different from those of a single densely ionizing beam, so that at low doses marked deviations from 1-beam action patterns may not occur. If, for mixtures of densely ionizing radiations in experiments on damaging endpoints, evidence against deviations from NSNA action accumulates, that will be more important, and more welcome, than would a more novel and exciting finding of major synergy for some specific mixture.

Given that synergy theory will continue to be emphasized, trying to find a systematic, quantitative, precisely defined approach to synergy theory that is general enough to cover most cases of interest and is self-consistent in the sense of obeying the sham mixture and the mixture of mixtures principles seems worthwhile. The indicated procedure is, we would suggest, to calculate NSNA results for each self-consistent synergy theory whose domain of validity includes the given mixture. There are currently only a few such candidate theories, and if none of them qualifies, SEA can serve as the last resort provided its weaknesses are explicitly taken into account. That procedure should serve the purposes of synergy calculations without inducing over-optimism about the reliability of the results.

**Acknowledgements** The authors thank Edward G. Huang for many useful discussions.

**Funding:** This work was supported in part by NASA Grant #80JSC021T9917 and by contract #DE-AC02-05CH11231 with the U.S. Department of Energy.

**Captions to figures**

**Fig. 1.** Interpreting the Loewe additivity NSNA EDR

The Loewe NSNA EDR shown gives a hypothetical calculated estimate of the NSNA mixture dose needed to produce a given deleterious effect in accordance with Eq. (1.3.2). If a mixture experiment were to find that a mixture dose smaller than the NSNA dose produces a given effect, that would indicate extra dose potency in the mixture – potency not expected from analyzing the action of individual agents, i.e. synergy. Hence the location in the dose-effect plane of the synergy and antagonism regions. As always in our context of analyzing damaging agents, synergy would count as "bad" synergy. Effect is here taken as the independent variable and dose as the dependent variable. Comments on the reason for, and the advantages of, such a switch are discussed in subsection 2.5. With only minor tweaks those comments apply equally to this figure and to the Loewe additivity synergy theory which it illustrates.

**Fig. 2.** EDR plot for quadratic and cubic DERs

The two mixture components, contributing equal doses, have EDRs $F_1(I) = I^{1/2}$ (blue curve) and $F_2(I) = I^{1/3}$ (black curve). The curves show the dose that would be needed by a mixture component if that component produced a given effect by itself. The mixture NSNA IEA EDR, see Eq. (2.4.13), is represented by the red curve. An observed mixture (dose, effect) pair below the red curve would indicate less dose than expected from experiments with the two components acting individually, i.e. synergy. Above the red curve is the antagonism region.

**Fig. 3.** 1-agent and synergy theory EDRs and DERs for a four-agent mixture

Panel A. 1-agent EDRs show the dose that would be needed by a mixture component if that component produced a given effect $E$ by itself. The SEA, Loewe additivity and IEA NSNA EDRs are calculated from the component DERs or EDRs using Eqs (1.3.1), (1.3.2) and (2.2.1) or (2.4.4), respectively. It is seen that the SEA NSNA EDR

lies beneath all four 1-agent EDRs in an interval centered near effect $E = 3$ and dose $d = 2$ exemplifying the fact that when the 1-agent DERs are highly nonlinear, as they are here, SEA should not be used.

For each synergy theory, an experimental mixture dose below the NSNA curve (black dot) would lie in the synergy region – less dose than expected from the component EDRs can produce a given effect – and this counts as "bad" synergy, since we deal with deleterious effects. In this paper, we did not discuss how far below an NSNA EDR curve a dose has to lie to qualify as statistically significant synergy – the answer to that question depends (among other things) on the statistical error structures of the 1-agent DERs after calibration from specific data. Such statistical metrics are beyond the scope of the paper. Thus, no error bars are shown for the dot.

The NSNA Loewe additivity curve lies between the lowest and highest 1-agent EDRs at each dose, exemplifying Proposition 1.3.1. Such "betweenness" is intuitively interpreted as saying that if there are no "extra" interactions a mixture effect should be in some sense an average of the component EDRs. The IEA curve here has this same betweenness property, and we will prove in subsection 3.2 that in general an IEA NSNA curve can never lie above the highest 1-agent EDR. Unfortunately, as will also be shown in subsection 3.2, it can sometimes lie below the lowest – even here, in panel A, it comes close to doing so in the same interval where SEA violates betweenness.

Panel B contains exactly the same information as panel A. Since an EDR is the inverse function of a DER, plotting DERs instead of EDRs merely involves a switching of horizontal and vertical axes, which induces diagonal reflection of curves.

**Fig. 4.** EDR plot illustrating the betweenness property in similarity theory

$F_1(I) = \exp(I/2)-1$ (blue curve) and $F_2(I) = \exp(I)-1$ (black curve) are EDRs of two equally contributing mixture components. The mixture EDR $F(I)$ is given in Eq. (2.4.4) and represented by the red curve. Note that, as in Case 1 discussed in subsection 3.1, the mixture EDR stays between the lower and upper individual EDR curves, i.e. has betweenness property. Therefore, the same also holds for the DERs.

**Fig. 5.** EDR plot illustrating the lack of betweenness

1-agent EDRs are $F_1(I) = I$ (blue line) and $F_2(I) = I^{1/2}$ (black curve) while the mixture NSNA EDR is represented by red curve. Near (effect, dose) point (1, 1) there is a two-dimensional region, above the red curve but below both the blue and the black curves, where a dose smaller than that needed by either of the two component agents produces the same effect. This region lies in the antagonism portion of figure, instead of being in the synergy portion. This betweenness violation contradicts the intuitive notion that a mixture dose more effective than any 1-component dose must be in the synergy region of the dose-effect plane. For more details, see Example 4 in subsection 3.2.

**Fig. 6.** Counterexample to eventual betweenness with two universally ordered piecewise linear EDRs

The red piecewise linear mixture EDR curve stays beneath the two universally ordered linear (or piecewise linear) component EDR curves in the effect interval (1.04, ∞). For more details, see Example 5 in subsection 3.2.

**Fig. 7a.** Mixture EDR for two universally ordered LQ DERs

1-agent LQ DERs are $E_1(d) = 2d + 2d^2$ (represented by blue EDR curve) and $E_2(d) = d + 0.5d^2$ (represented by black EDR curve). Mixture EDR is shown as the red curve. Note the betweenness property.

**Fig. 7b.** Mixture EDR for two intersecting individual LQ DERs

1-agent LQ DERs are $E_1(d) = 5d+d^2$ (represented by blue EDR curve) and $E_2(d) = d+4d^2$ (represented by black EDR curve). Mixture EDR is shown as the red curve. The betweenness property is violated near the intersection point.

**Fig. 8a.** Mixture EDR for two DERs with distinct (one finite and one infinite) ranges of effect

Component EDRs are $F_1(I) = I$ (blue line) and $F_2(I) = 3I/(1-I)$ (black curve). The mixture EDR (red curve) is given by Eq. (3.3.1a). Notice the betweenness property.

**Fig. 8b.** Mixture EDR for two DERs with distinct (one finite and one infinite) ranges of effect

Component EDRs are $F_1(I) = I$ (blue line) and $F_2(I) = I/[3(1-I)]$ (black curve). Mixture EDR (red curve) is given by Eq. (3.3.1b). A violation of betweenness occurs near the intersection point.

**Fig. 9a.** Mixture EDR for two DERs with distinct finite ranges of effect that has the betweenness property

1-agent EDRs are $F_1(I) = I/(1-I)$ (blue curve) and $F_2(I) = I/(2-I)$ (black curve). The mixture EDR (red curve) is given by Eq. (3.3.2) with $a = b = 1$.

**Fig. 9b.** Mixture EDR for two DERs with distinct finite ranges of effect that violates the betweenness property

1-agent EDRs are $F_1(I) = I/(1-I)$ (blue curve) and $F_2(I) = I/(2-I)$ (black curve). The mixture EDR (red curve) is given by Eq. (3.3.2) with $a = b = 12$.

**Fig. 10.** Dose mixture of two agents having DERs with distinct (one finite and one infinite) ranges of effect may violate the betweenness property but always has the eventual betweenness property

$F_1(I)$ (blue curve) is given by Eq. (3.3.3) and $F_2(I)$ (black curve) is given by Eq. (3.2.3). The 1:1 mixture NSNA EDR is given by Eq. (3.3.4) (red curve). In the effect interval $(11/9, I^*) \sim (1.222, 1.308)$ the red curve lies below the remaining 1-agent DER, but on the effect interval $(I^*, \infty)$, betweenness holds in the sense that the red curve stays above $F_2(I)$.

**Fig. 11.** Mixture NSNA EDR for four agents with different (one finite and three infinite) ranges of effect

EDRs of the four agents are $F_1(I) = 1/\ln(1/I)$ (blue curve), $F_2(I) = \exp(I)-1$ (black curve), $F_3(I) = I^2$ (magenta curve) and $F_4(I) = (2I+1)^{1/2}-1$ (green curve). Note that although betweenness holds, the mixture EDR (red curve) is surprisingly low, a signal that even though an experiment may be showing synergy, the NSNA EDR may be pointing to antagonism instead. The mixture EDR is given by Eqs (3.3.5a) and (3.3.5b).

**Fig. 1**

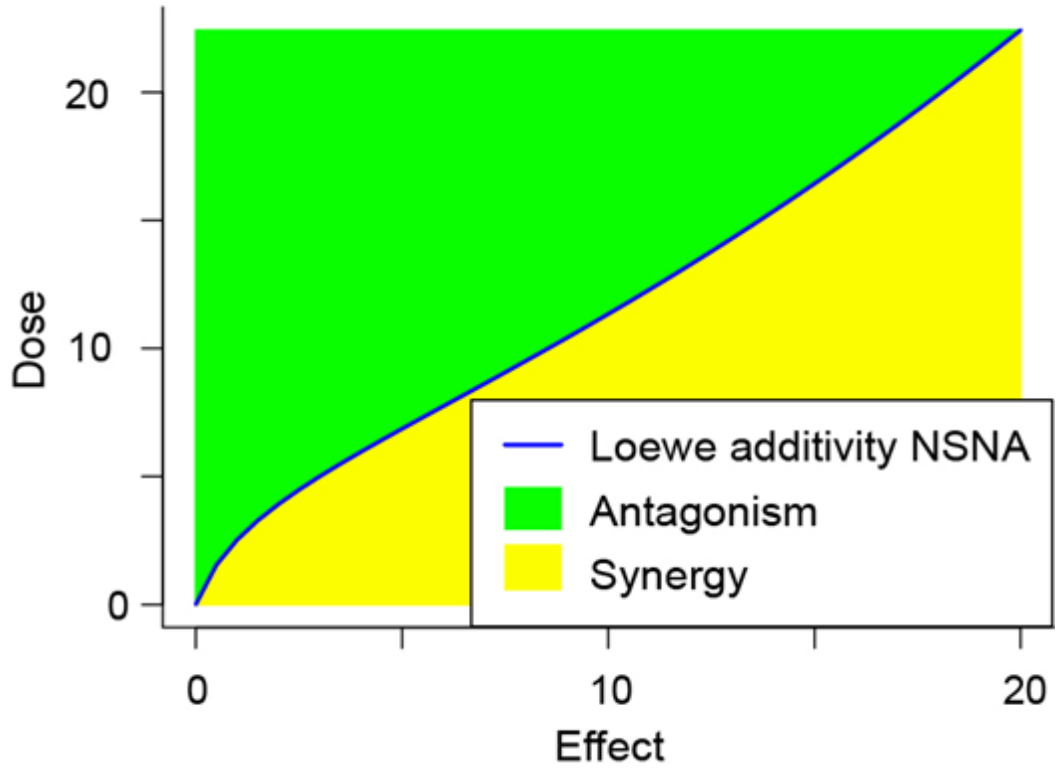

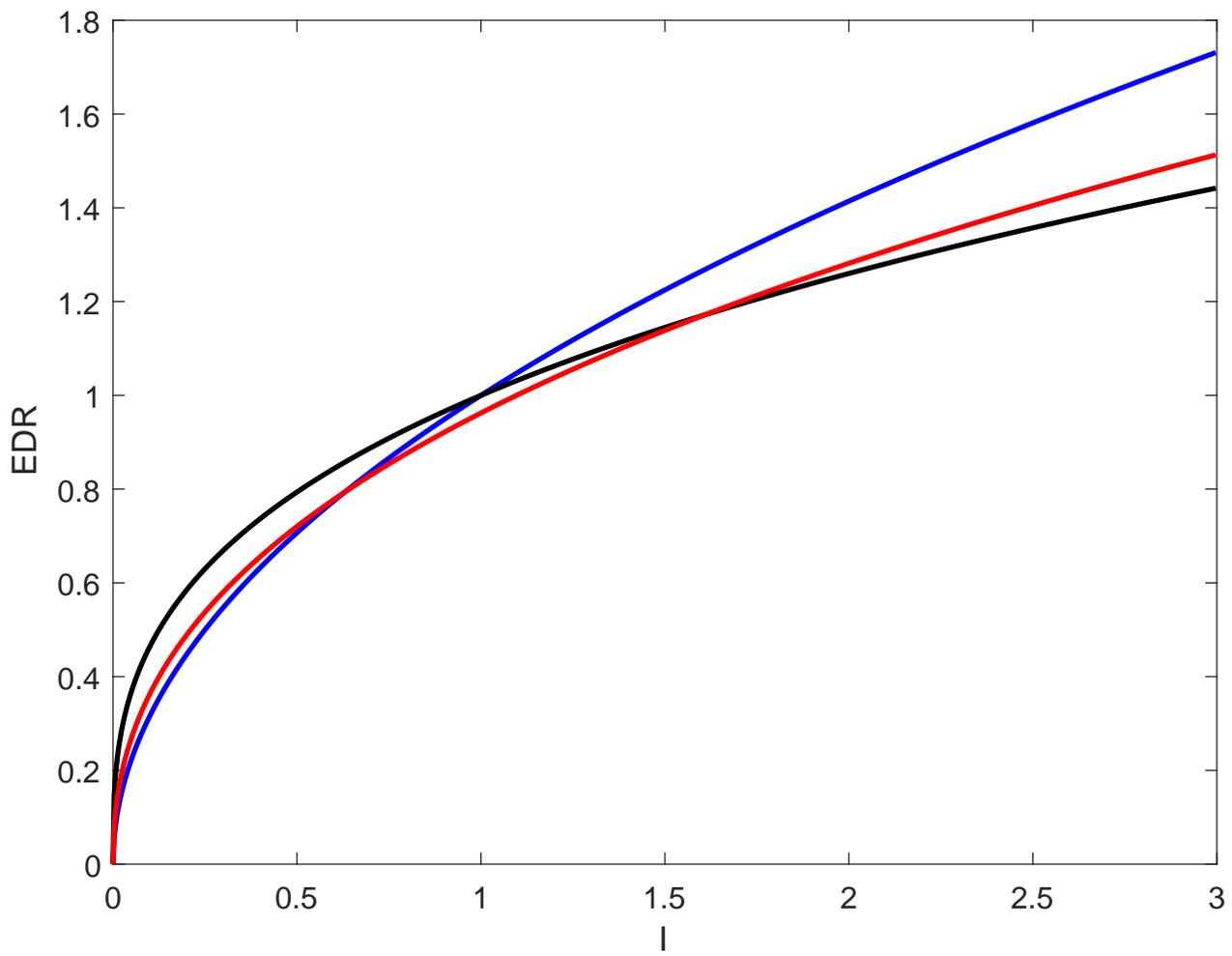

Fig. 2

**Fig. 3**

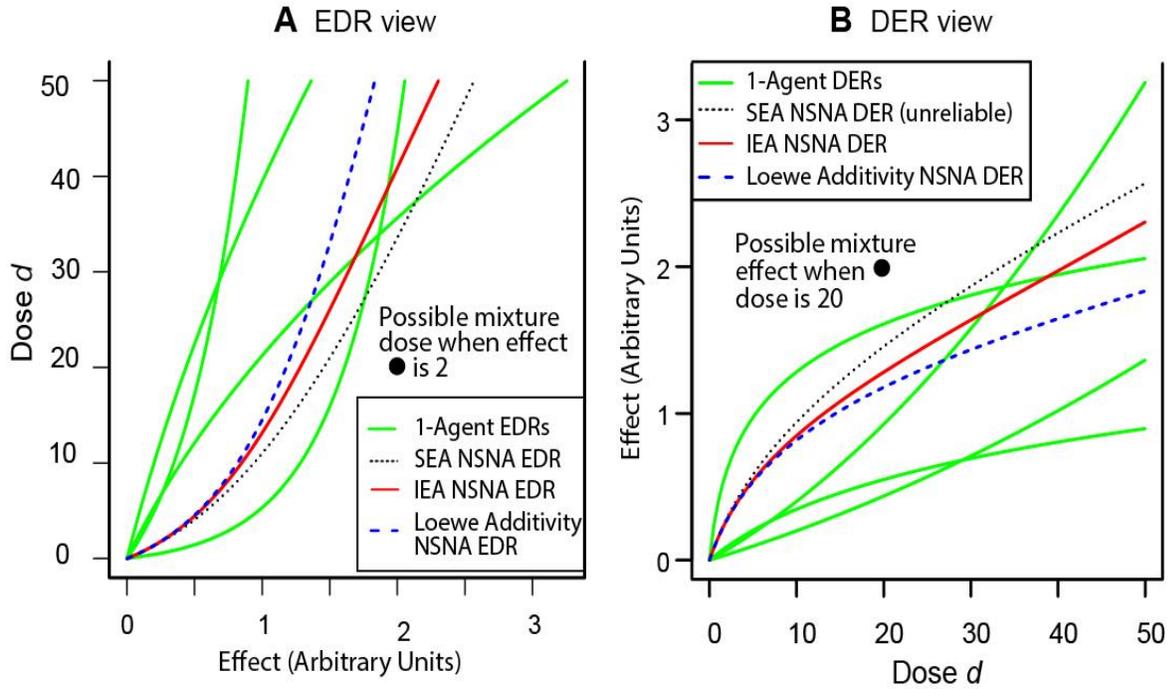

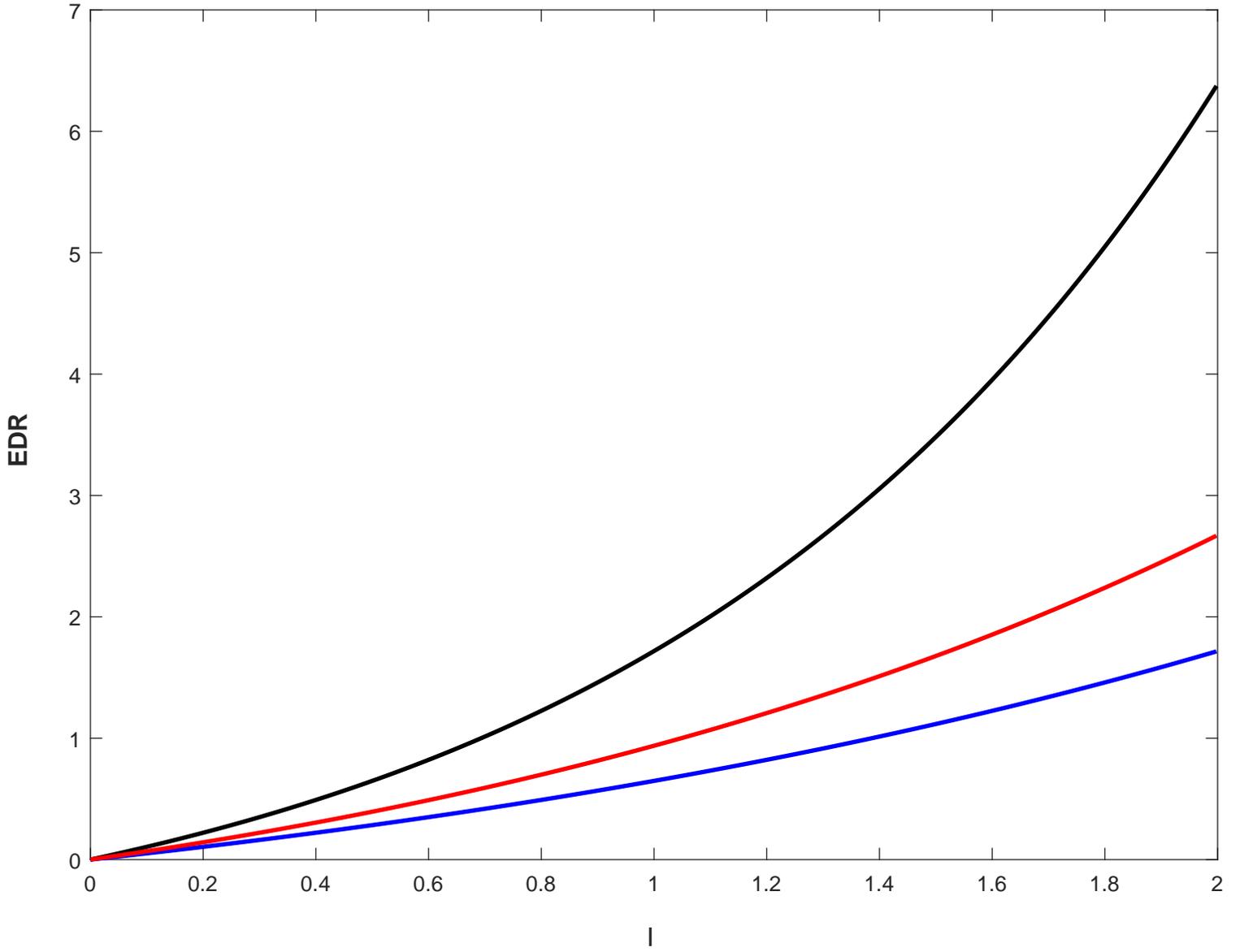

**Fig. 4**

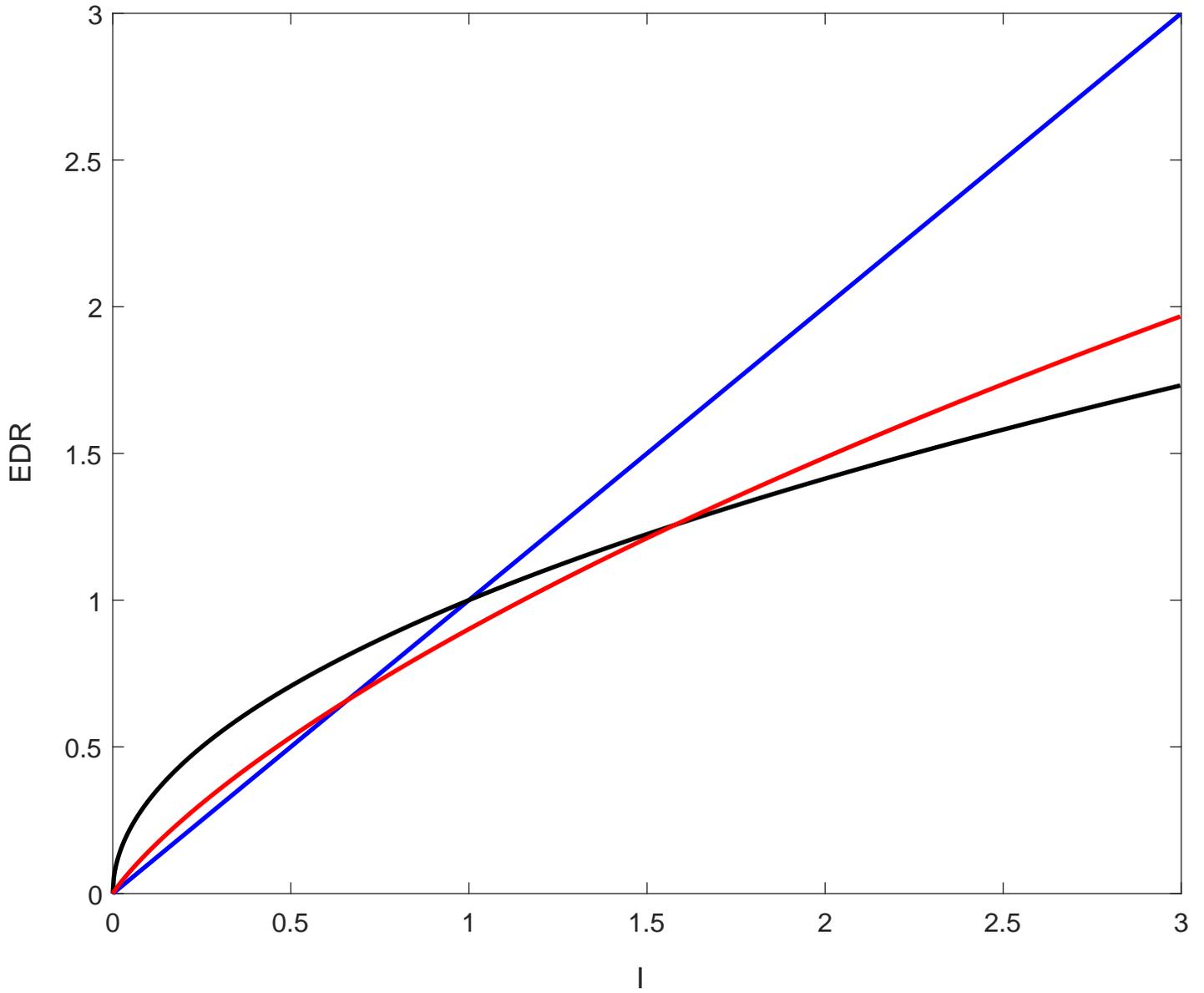

Fig. 5

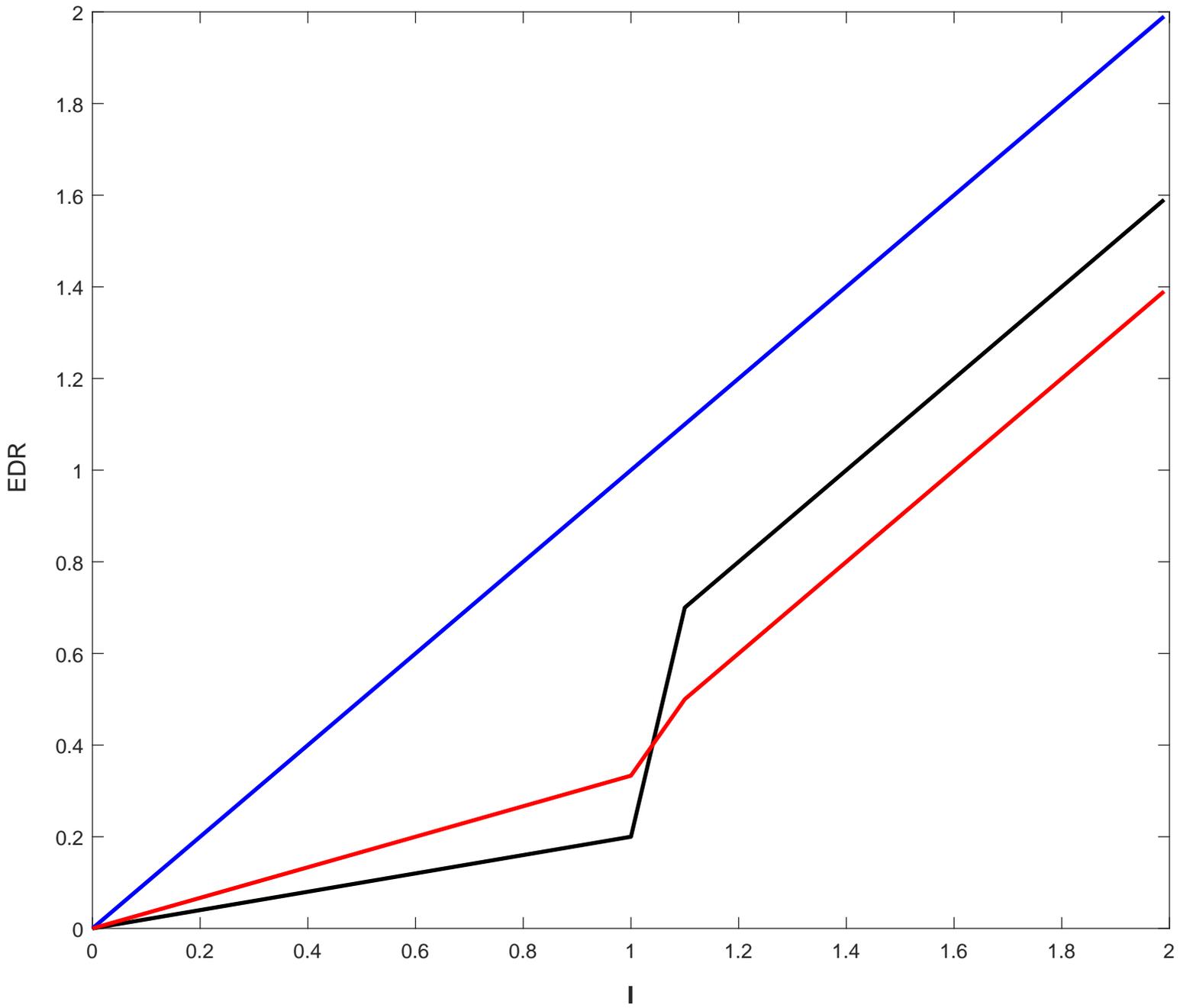

Fig. 6

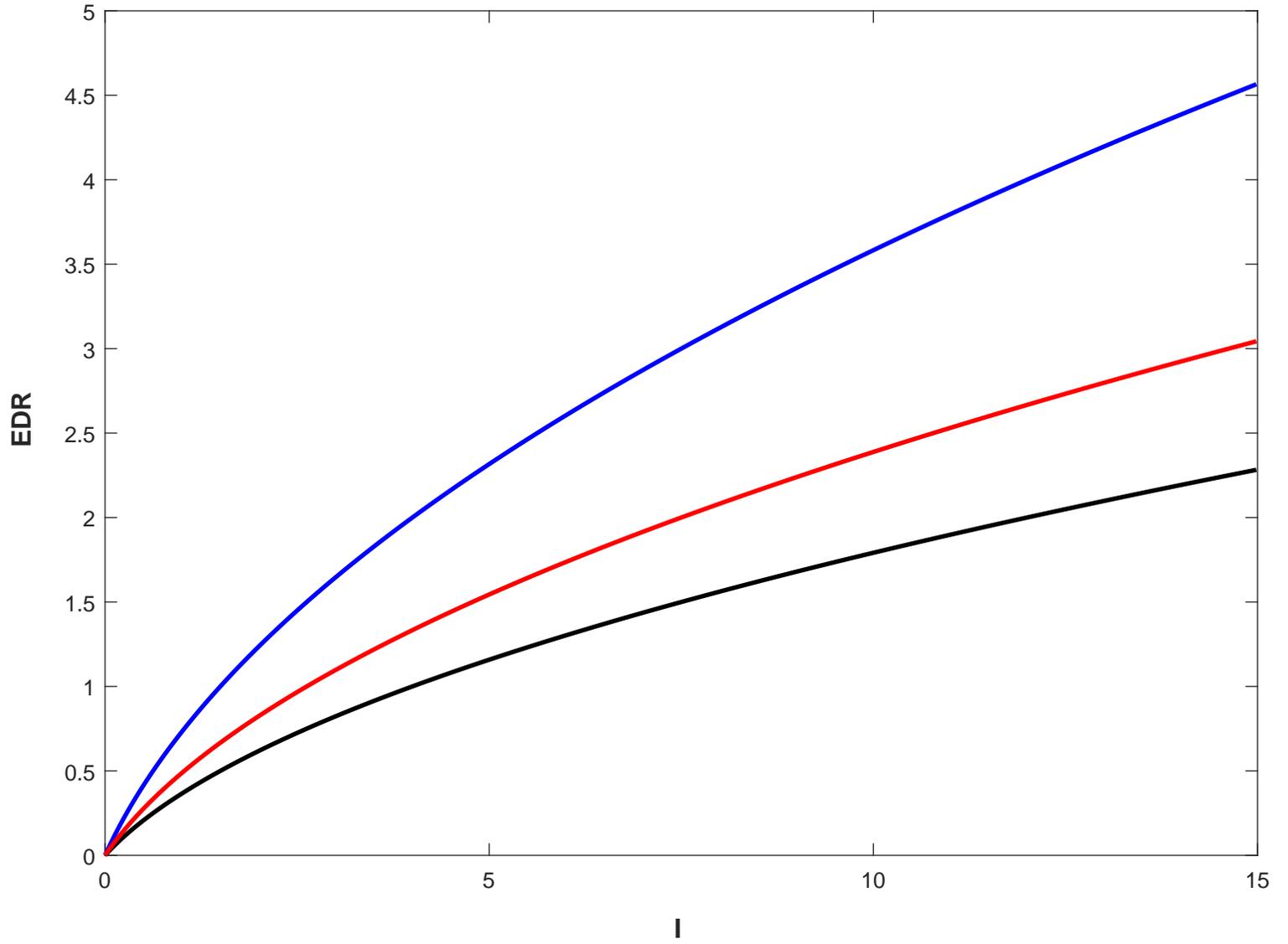

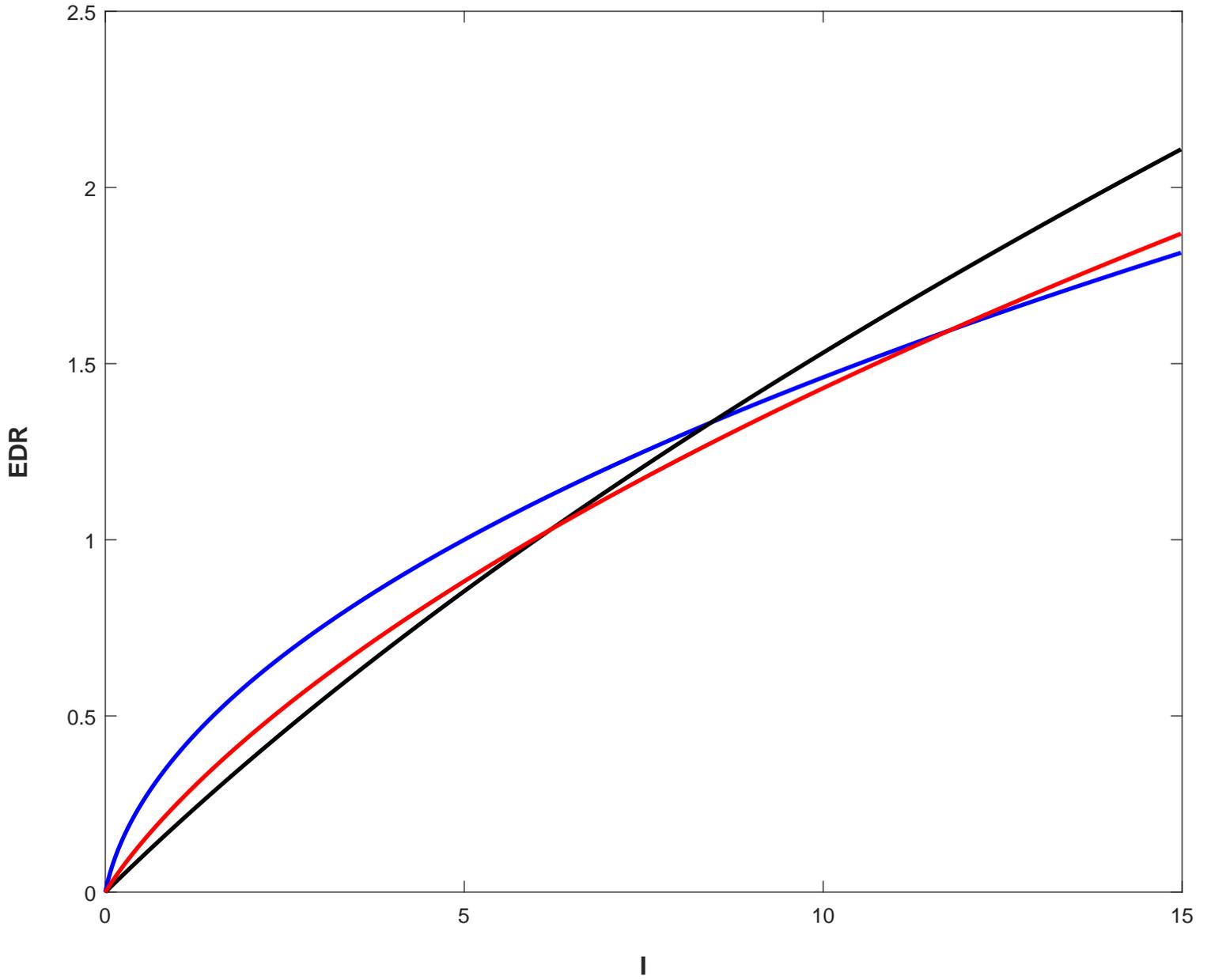

**Fig. 7b**

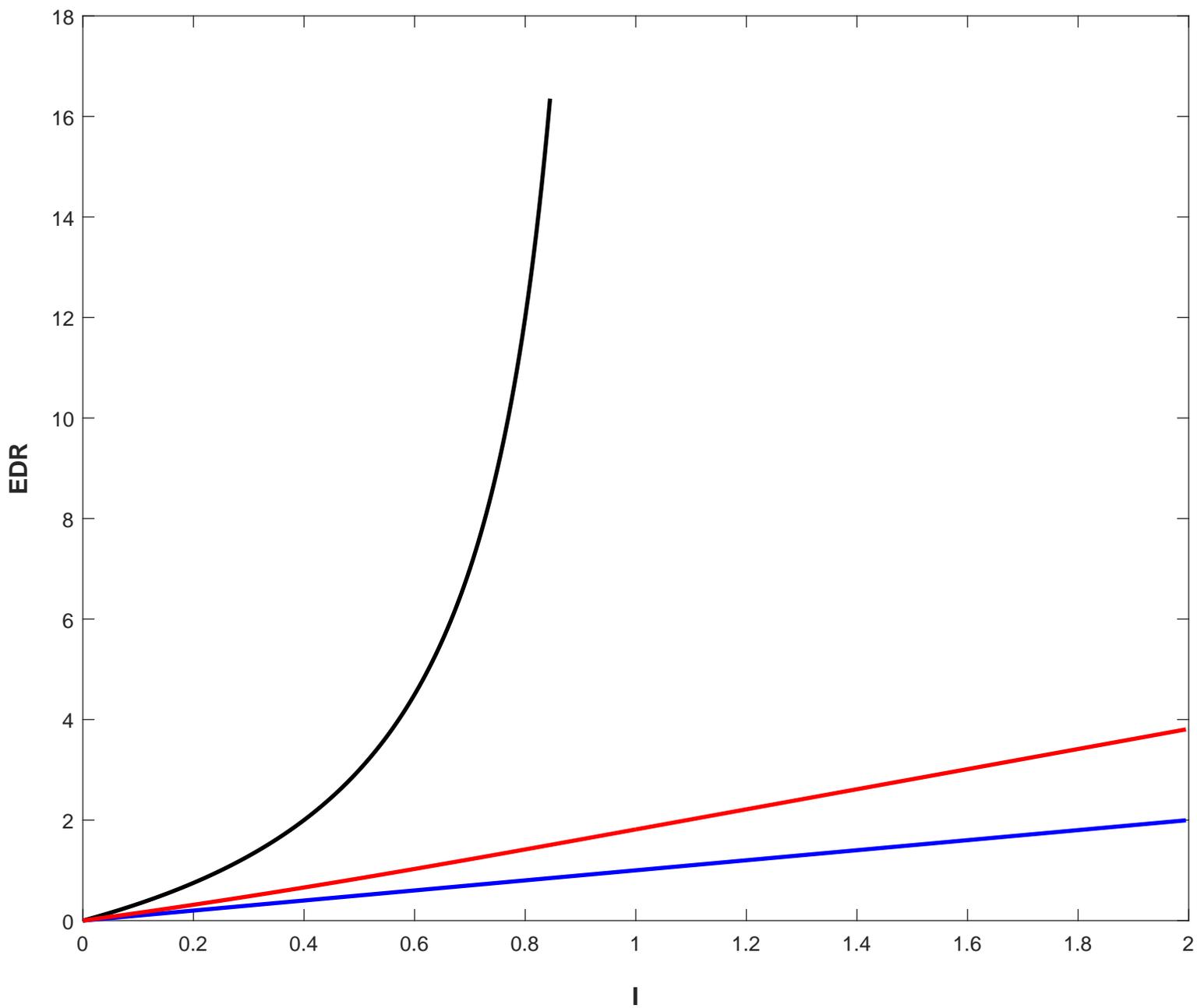

Fig. 8a

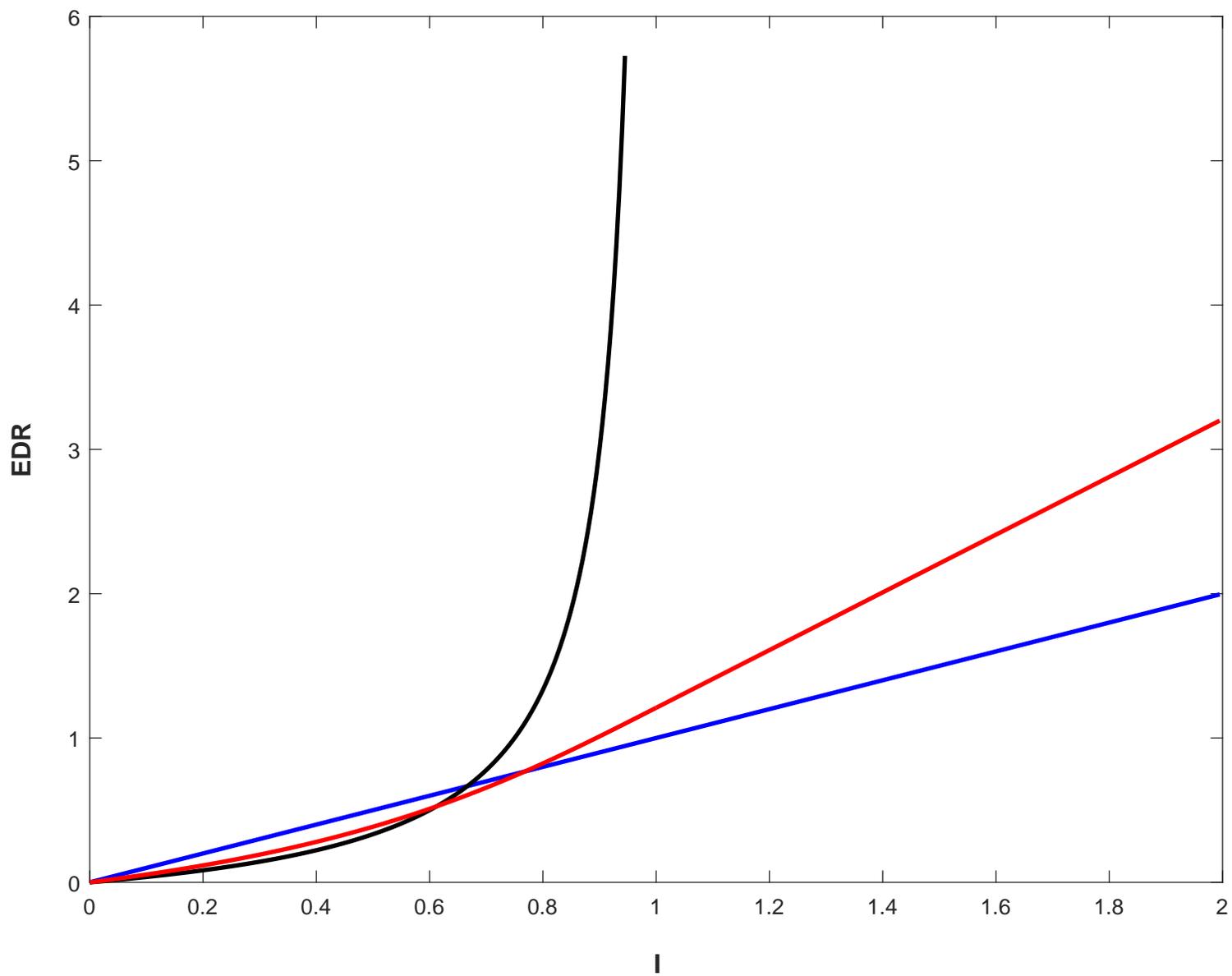

Fig. 8b

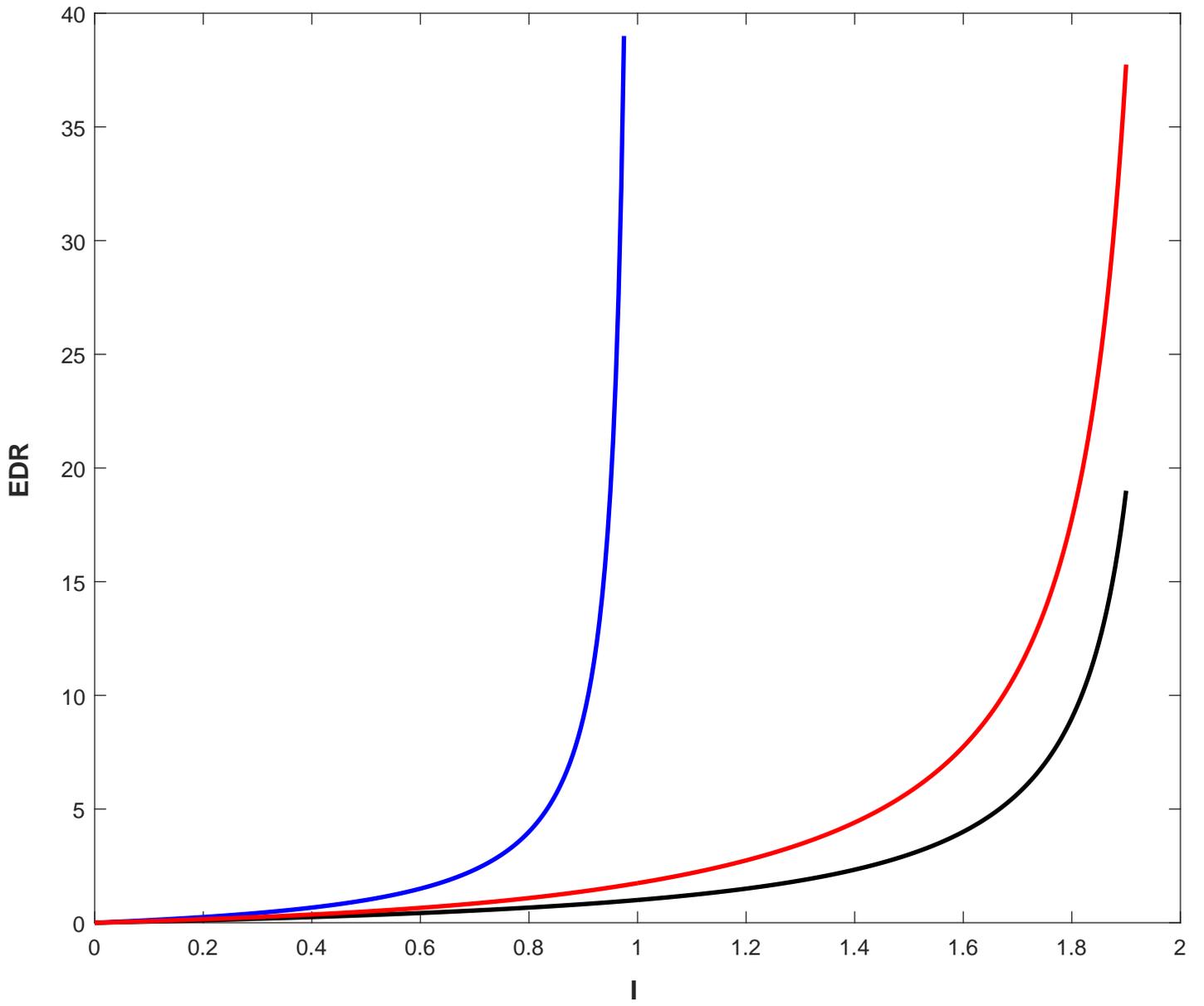

Fig. 9a

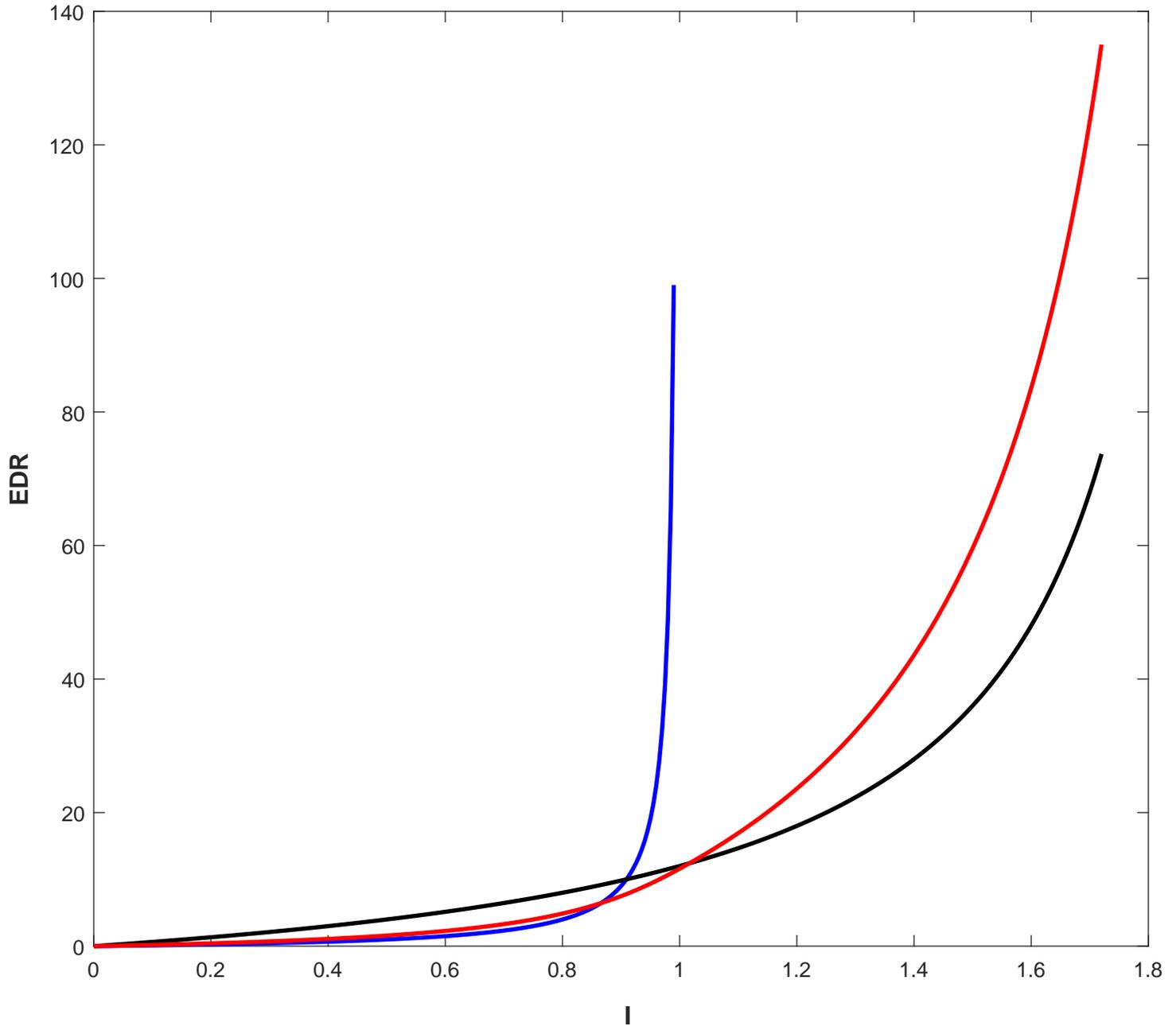

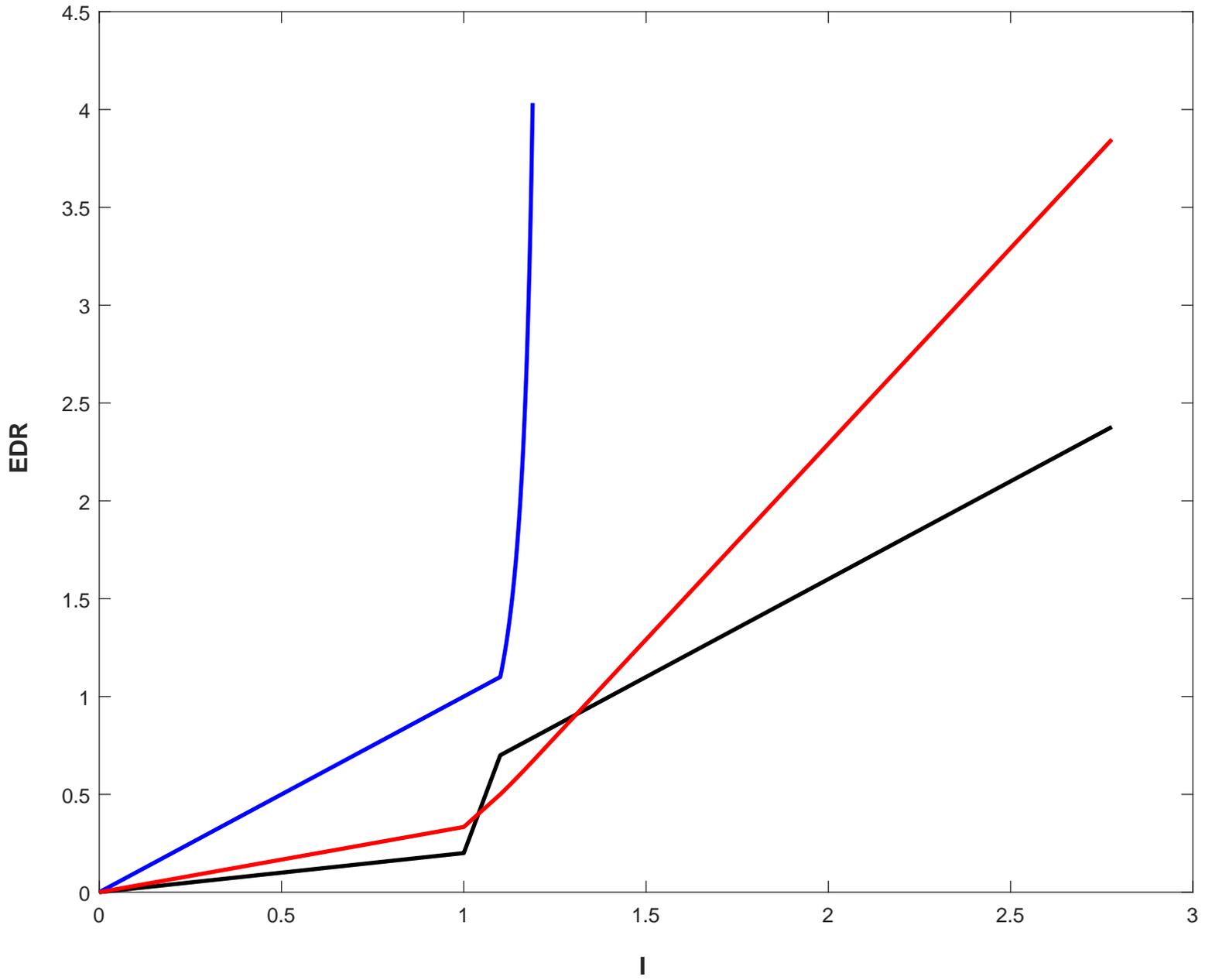
Fig. 10

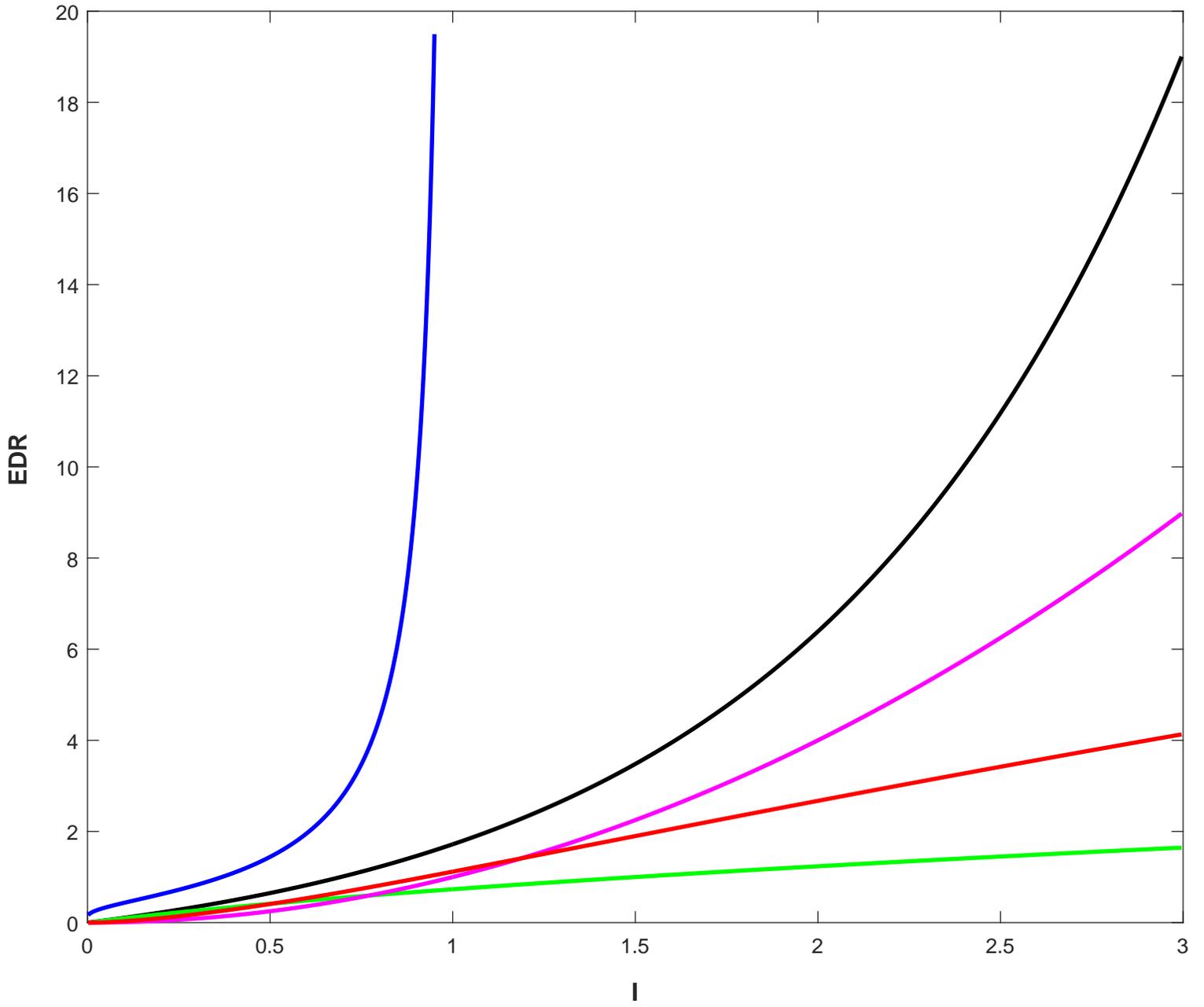

**Fig. 11**

# Supplementary Materials

## R Script for Fig. 1

```r
# This is the script illustrating Loewe interpretation. Copyright GPLv3.
# Use freely but at your own risk, no warranty given or implied
# This is the script for a 4-component mixture. Copyright GPLv3.
# Written by Liyang Xie and Rainer Sachs, first half of 2021
# Range of effects
effect_range <- .5*0:40
# Scale factors
a1 <- 2
a2 <- 0.15
# The NSNA curve
d <- 2*log(1+a1*effect_range)+(a2 * effect_range + 1)^2 - 1
# Plot the NSNA
plot(c(0,20),c(0,d[41]),col='black',ylab='dose',xlab='Effect',bty='l')
polygon(c(effect_range,20,0), c(d,d[41],d[41]), col= 'green', border='green')
polygon(c(effect_range,20,0), c(d,0,0), col= 'yellow', border='yellow')
lines(effect_range,d,type ='l',col='blue',lwd=2)
# legend(x=9.5, y=7,
#      legend = c("Loewe additivity NSNA",","antagonism",", "synergy"),
#      col = c("blue",'black', "green",'black', "yellow"),
#      lwd = c(2,1,20,1,20), cex=0.85)
legend(x=2.5, y=3.5,col='black',legend = c("Synergy Region"),cex=.95, lty=1)
legend(x=.1,y=21.3,col='black',legend = c("Antagonism  Region"),cex  =  .95)
plot(c(0,20),c(0,d[41]),col='black',ylab='dose',xlab='Effect',bty='l')
polygon(c(effect_range,20,0), c(d,d[41],d[41]), col= 'gray', border='black',density =20, angle = 0)
polygon(c(effect_range,20,0), c(d,0,0), border='black', density =20, angle = 90, col='gray')
lines(effect_range,d,type ='l',col='black', lwd=2)
legend(x=5, y=4,col='black',legend = c("Synergy Region"),cex=.95)
legend(x=.1,y=21.3,col='black',legend = c("Antagonism  Region"),cex  =  .95)
```

## Matlab Script for Fig. 2

```matlab
clear all
M=3;
n=1000;
for k=1:n
```

```
    x(k)=M*(k-1)/n;
    u(k)=sqrt(x(k));
    y(k)=x(k)^(1/3);
    v(k)=sqrt(y(k));
    z(k)=2*y(k)-8*v(k)/3+16*log(1+1.5*v(k))/9;
end
plot(x,u,'b','Linewidth',2)
hold on
plot(x,y,'k','Linewidth',2)
hold on
plot(x,z,'r','Linewidth',2)
hold off
```

## R Script for Fig. 3

```
# This is the script for a 4-component mixture. Copyright GPLv3.
# Written by Liyang Xie, January 2021. Modifications by Rainer Sachs, later 2021
# Use and modify freely but at your own risk, no warranty given or implied
rm(list=ls())
# Use the deSolve package for solving ODEs
library(deSolve)
d = 0.0001*0:100; E=0.0004*0:100 # dose and effect ranges
K1=200;a1=1000000;K2=10;a2=20;K3=2000;a3=2000;K4=1/8;a4=1;
a4=.5;K4= 1/4;a3=500;K3 =500; a1=100000 # these values make
# SEA violate betweeness & IEA obey it. The other values
# above make both synergy theoriea -- SEA & IEA -- violate it.
E1 = log(1+a1*d)/K1 # generator (a1/K1)*exp(-K1*E)
d1= (exp(K1*E) -1)/a1
E2 = ((a2*d+1)^(1/2) -1)/K2 # generator (a2/K2)/(2+2*K2*E)
d2 =(2*K2*E+(K2*E)^2)/a2
E3=((a3*d+1)^2 - 1)/K3 # generator (a3/K3)(1+K3*E)^(1/2)
d3 = ((K3*E+ 1)^(1/2)- 1)/a3
E4 =(exp(a4*d) -1)/K4 # generator K4*E(a4/K4)*(1+K4*E)
d4 = log(1+K4*E)/a4
r1=1/4;r2=1/4;r3=1/4;r4=1/4; dsea= r1*d # all 4 agents contribute
# equaldoses to mIxture, each contributes dsea
params <- c(K1,a1,r1,K2,a2,r2,K3,a3,r3,K4,a4,r4)
# SEA adds up each component's effect
SEA=log(1+a1*dsea)/K1 +((a2*dsea+1)^(1/2) -1)/K2 +
```

```
   ((a3*dsea+1)^2 - 1)/K3 + (exp(a4*dsea) -1)/K4
# Returns the derivative at dose d and effect I
dI <- function(d, I, params) {
    G1 <- (a1/K1)*exp(-K1*I)
    G2 <- (a2/K2)/(2+2*K2*I)
    G3 <- (a3/K3)*(1+K3*I)^(1/2)
    G4 <- (a4/K4)*(1+K4*I)
    return(list(r1*G1 + r2*G2+ r3*G3 + r4*G4))
}
# Solve the ODE to get the IEA mixture DER
IEA <- ode(c(I = 0), d, dI, params) # Next plot DERs
plot(d,E4,type ='l',col='green',xlab='dose',ylab='',bty='u',ylim= c(0,.04))
lines(d, E1, col ='green')
lines(d, E2, col ='green')
lines(d, E3, col = 'green')
lines(d, SEA, col = 'black', lty = 3)
lines(d, IEA[,'I'], col = 'red')
# Loewe additivity NSNA mixture dose is calculated as
# 1/d = sum of rj / dj(E). where dj is the jth EDR
# Effect range [0,0.03], approximates dose range [0, 0.01]
Loewe_Add_d <- 1/((r1/d1)+(r2/d2)+(r3/d3)+(r4/d4))
#lines(Loewe_Add_d, E, col = 'blue', lty = 2) #RKS: this does not work
# Now plot EDRs & Loewe NSNA curve.
plot(E,d1,type ='l',col='green',xlab= 'E',ylab='',bty='u', ylim=c(0, .01))
lines(E,d2, col = 'green')
lines(E,d3,col = 'green')
lines(E,d4, col = 'green')
#lines(SEA, d,col = 'black', lty = 3)#RKS: this and the next line do not work
lines(IEA[,'I'],IEA[,'time'], col = 'red')
lines(E, Loewe_Add_d, col = 'blue', lty = 2)
# legend("bottomright", legend = c("1-agent EDRs", "SEA", "IEA", "Loewe Additivity"),
#     col = c("green", "black", "red", "blue"), lty = c(1,3,1,2), cex=0.85)
# at very end, rescale dose and effect axes in plots already exported to your
# graphics program so as to avoid decimal points.
```

## Matlab Script for Fig. 4

```
clear all
M=2;
```

```
n=1000;
a=log(3)-2;
for k=1:n
   x(k)=M*(k-1)/n;
   u(k)=exp(x(k)/2);
   y(k)=u(k)-1;
   z(k)=exp(x(k))-1;
   w(k)=2*u(k)-log(2*u(k)+1)+a;
end
plot(x,y,'b','Linewidth',2)
hold on
plot(x,z,'k','Linewidth',2)
hold on
plot(x,w,'r','Linewidth',2)
hold off
```

**Matlab Script for Fig. 5**

```
clear all
M=3;
n=1000;
for k=1:n
   x(k)=M*(k-1)/n;
   y(k)=sqrt(x(k));
   z(k)=2*y(k)-log(1+2*y(k));
end
plot(x,x,'b','Linewidth',2)
hold on
plot(x,y,'k','Linewidth',2)
hold on
plot(x,z,'r','Linewidth',2)
hold off
```

**Matlab Script for Fig. 6**

```
clear all
for k=1:100
   x(k)=(k-1)/100;
   y(k)=x(k)/5;
   z(k)=x(k)/3;
```

```matlab
end
for k=101:111
    x(k)=(k-1)/100;
    y(k)=5*x(k)-4.8;
    z(k)=5*x(k)/3-4/3;
end
for k=112:200
    x(k)=(k-1)/100;
    y(k)=x(k)-0.4;
    z(k)=x(k)-0.6;
end
plot(x,x,'b','Linewidth',2)
hold on
plot(x,y,'k','Linewidth',2)
hold on
plot(x,z,'r','Linewidth',2)
hold off
```

**Matlab Script for Fig. 7a**

```matlab
clear all
M=15;
n=1000;
a1=1; b1=0.5;
a2=2; b2=2;
F=@(x)1./(sqrt(a1*a1+4*b1*x)+sqrt(a2*a2+4*b2*x));
for k=1:n
    u(k)=M*(k-1)/n;
    y(k)=(sqrt(a1*a1+4*b1*u(k))-a1)/(2*b1);
    z(k)=(sqrt(a2*a2+4*b2*u(k))-a2)/(2*b2);
    w(k)=2*quad(F,0,u(k));
end
plot(u,y,'b','Linewidth',2)
hold on
plot(u,z,'k','Linewidth',2)
hold on
plot(u,w,'r','Linewidth',2)
hold off
```

**Matlab Script for Fig. 7b**

```
clear all
M=15;
n=1000;
a1=1; b1=4;
a2=5; b2=1;
F=@(x)1./(sqrt(a1*a1+4*b1*x)+sqrt(a2*a2+4*b2*x));
for k=1:n
    u(k)=M*(k-1)/n;
    y(k)=(sqrt(a1*a1+4*b1*u(k))-a1)/(2*b1);
    z(k)=(sqrt(a2*a2+4*b2*u(k))-a2)/(2*b2);
    w(k)=2*quad(F,0,u(k));
end
plot(u,y,'b','Linewidth',2)
hold on
plot(u,z,'k','Linewidth',2)
hold on
plot(u,w,'r','Linewidth',2)
hold off
```

**Matlab Script for Fig. 8a**

```
clear all
a=sqrt(3);
for k=1:400
    t(k)=(k-1)/200;
end
for i=1:170
    s(i)=(i-1)/200;
    y(i)=3*s(i)/(1-s(i));
end
for j=1:201
    v(j)=(j-1)/200;
    z(j)=pi/a-2*a*atan((1-v(j))/a);
end
for l=1:200
    u(l)=1+(l-1)/200;
    w(l)=pi/a+2*(u(l)-1);
end
```

```
plot(t,t,'b','Linewidth',2)
hold on
plot(s,y,'k','Linewidth',2)
hold on
plot(v,z,'r','Linewidth',2)
hold on
plot(u,w,'r','Linewidth',2)
hold off
```

## Matlab Script for Fig. 8b

```
clear all
a=sqrt(3);
for k=1:400
    t(k)=(k-1)/200;
end
for i=1:190
    s(i)=(i-1)/200;
    y(i)=s(i)/(3*(1-s(i)));
end
for j=1:201
    v(j)=(j-1)/200;
    z(j)=2*(pi/3-atan(a*(1-v(j))))/a;
end
for l=1:200
    u(l)=1+(l-1)/200;
    w(l)=2*pi/(3*a)+2*(u(l)-1);
end
plot(t,t,'b','Linewidth',2)
hold on
plot(s,y,'k','Linewidth',2)
hold on
plot(v,z,'r','Linewidth',2)
hold on
plot(u,w,'r','Linewidth',2)
hold off
```

## Matlab Script for Fig. 9a

```
clear all
```

```
a=1; b=1;
C=2*sqrt(2)*(atan(2*sqrt(2))-atan(sqrt(2)/2));
for k=1:196
   s(k)=(k-1)/200;
   x(k)=s(k)/(1-s(k));
end
for j=1:201
   t(j)=(j-1)/200;
   y(j)=t(j)/(2-t(j));
   z(j)=2*sqrt(2)*(atan(2*sqrt(2))+atan(1.5*sqrt(2)*t(j)-2*sqrt(2)));
end
for i=1:181
   u(i)=1+(i-1)/200;
   v(i)=u(i)/(2-u(i));
   w(i)=C+4*(u(i)-1)/(2-u(i));
end
plot(s,x,'b','Linewidth',2)
hold on
plot(t,y,'k','Linewidth',2)
hold on
plot(t,z,'r','Linewidth',2)
hold on
plot(u,v,'k','Linewidth',2)
hold on
plot(u,w,'r','Linewidth',2)
hold off
```

**Matlab Script for Fig. 9b**

```
clear all
a=1; b=12;
C=2*sqrt(2*a*b)*(atan(2*(a+b)/sqrt(2*a*b))-atan(sqrt(a/(2*b))));
for k=1:199
   s(k)=(k-1)/200;
   x(k)=a*s(k)/(1-s(k));
end
for j=1:201
   t(j)=(j-1)/200;
   y(j)=b*t(j)/(2-t(j));
```

```
    z(j)=2*sqrt(2*a*b)*(atan(2*(a+b)/sqrt(2*a*b))+atan((a+2*b)*t(j)/sqrt(2*a*b)-2*(a+b)/sqrt(2*a*b)));
end
for i=1:145
    u(i)=(1+(i-1)/200);
    v(i)=b*u(i)/(2-u(i));
    w(i)=C+4*b*(u(i)-1)/(2-u(i));
end
plot(s,x,'b','Linewidth',2)
hold on
plot(t,y,'k','Linewidth',2)
hold on
plot(t,z,'r','Linewidth',2)
hold on
plot(u,v,'k','Linewidth',2)
hold on
plot(u,w,'r','Linewidth',2)
hold off
```

**Matlab Script for Fig. 10**

```
clear all
for k=1:900
    x(k)=(k-1)/900;
    y(k)=x(k)/5;
    z(k)=x(k)/3;
end
for k=901:991
    x(k)=(k-1)/900;
    y(k)=5*x(k)-4.8;
    z(k)=5*x(k)/3-4/3;
end
for k=992:1100
    x(k)=(k-1)/900;
    y(k)=x(k)-0.4;
    z(k)=0.5+11*(atan(1/3)-atan(10/3-30*x(k)/11))/15;
end
for k=1101:2501
    x(k)=(k-1)/900;
    y(k)=x(k)-0.4;
```

```matlab
    z(k)=2*x(k)-35/18+11*atan(1/3)/15;
end
for i=1:991
   u(i)=(i-1)/900;
   v(i)=u(i);
end
for i=992:1071
   u(i)=(i-1)/900;
   v(i)=1.21/(11-9*u(i));
end
plot(u,v,'b','Linewidth',2)
hold on
plot(x,y,'k','Linewidth',2)
hold on
plot(x,z,'r','Linewidth',2)
hold off
```

**Matlab Script for Fig. 11**

```matlab
clear all
F=@(x)x./(1+2*x.*x.*log(x).*log(x)+2*x.*sqrt(2*x+1)+2*x.*exp(-x));
G=@(x)x./(1+2*x.*sqrt(2*x+1)+2*x.*exp(-x));
n=500;
for i=1:475
   u(i)=i/n;
   x(i)=-1/log(u(i));
end
for j=1:n
   v(j)=j/n;
   y1(j)=exp(v(j))-1;
   z1(j)=sqrt(2*v(j)+1)-1;
   w1(j)=v(j)*v(j);
   m1(j)=8*quad(F,0,v(j));
end
for k=1:n
   t(k)=1+2*(k-1)/n;
   y2(k)=exp(t(k))-1;
   z2(k)=sqrt(2*t(k)+1)-1;
   w2(k)=t(k)*t(k);
```

```matlab
    m2(k)=m1(n)+8*quad(G,1,t(k));
end
plot(u,x,'b','Linewidth',2)
hold on
plot(v,y1,'k','Linewidth',2)
hold on
plot(v,z1,'g','Linewidth',2)
hold on
plot(v,w1,'m','Linewidth',2)
hold on
plot(v,m1,'r','Linewidth',2)
hold on
plot(t,y2,'k','Linewidth',2)
hold on
plot(t,z2,'g','Linewidth',2)
hold on
plot(t,w2,'m','Linewidth',2)
hold on
plot(t,m2,'r','Linewidth',2)
hold off
```